\documentclass[preprint,12pt,3p]{elsarticle}

\usepackage{amssymb}
\usepackage{amsmath}
\usepackage{amsthm}
\usepackage{booktabs}
\usepackage{graphicx}
\usepackage{subfigure}
 \usepackage{float}
\newtheorem{definition}{Definition}[section]
\newtheorem{proposition}{Proposition}[section]
\journal{Communications in Nonlinear Science and Numerical Simulation}

\begin{document}

\begin{frontmatter}

\title{A soft Lasso model for the motion of a ball falling in the non-Newtonian fluid}
%\tnotetext[label0]{This is only an example}

\author[label1,label2]{Zongmin Wu\corref{cor1}%\fnref{label3}
}

\address[label1]{School of Mathematical Sciences, Fudan University, China}%\fnref{label4}
\address[label2]{School of Big Data and Statistics, Anhui University, China}

%\cortext[cor1]{I am corresponding author}
%\fntext[label3]{I also want to inform about\ldots}
%\fntext[label4]{Small city}

\ead{zmwu@fudan.edu.cn}
%\ead[url]{author-one-homepage.com}

\author[label1]{Ran Yang}

\ead{ryang19@fudan.edu.cn}

\begin{abstract}
From the mesoscopic point of view, a new concept of soft matching for mass points is proposed.
Then a soft Lasso's approach to learn the soft dynamical equation for the physical mechanical relationship is proposed, too.
Furthermore, a discrete iterative algorithm combining the Newton-Stokes term and the
soft Lasso’s term is developed to simulate the  motion of a ball falling in non-Newtonian fluids. The theory is validated by numerical examples and shows satisfactory results, which
exhibit the chaotic phenomena, sudden accelerations and the continual random oscillations.  The pattern of the motion is independent of initial values  and is preserved for long time.
\end{abstract}


\end{frontmatter}

% \linenumbers

\section{Introduction}
The problem of falling ball stems from the Galileo's free body falling, that is, ``the distance is proportional to the square of time''. This helps Newton to establish the second law $f=ma$ in classical mechanics  which means the force is equal to the mass multiplying the acceleration.  It can be expressed by an ordinary differential equation (ODE) that the derivative of velocity is equal to a constant $v'=c$. The constant $c$ is just equal to the acceleration of gravity $g$ in the motion of free falling body. The solutions of the ODE can be written as $v=ct+v_0$, where $v_0$ is the initial velocity. In this paper, we take the initial time as $0$. We can discretize the ODE $v'=c$  to get a discrete iterative algorithm $v_{n+1}-v_{n}=c(t_{n+1}-t_{n})$ to simulate the linear inertial motion.
\par
Based on Archimedes' floating principle, Stokes takes the resistance caused by the fluid into account that the resistance is proportional to the volumes of the fluid pushed away by a falling ball \cite{stokes}.
Then an ODE in the form of $v'=c-\lambda v$ is obtained to describe the motion of a falling ball. The ball will eventually approach to a terminal velocity $v_T$  and the solutions of this ODE can be written as  $v=v_T -(v_T-v_0)e^{-\lambda t}$  satisfying  $\lambda v_T=c$.
We can also discretize this ODE  to get an iterative algorithm $v_{n+1}-v_{n-1}=(c-\lambda v_n)(t_{n+1}-t_{n-1})$ to simulate the velocity of the motion.
\par
\begin{figure}[h]
		\centering
%\vspace{-2cm}
		\includegraphics[width=3.5in, height = 2.5in]{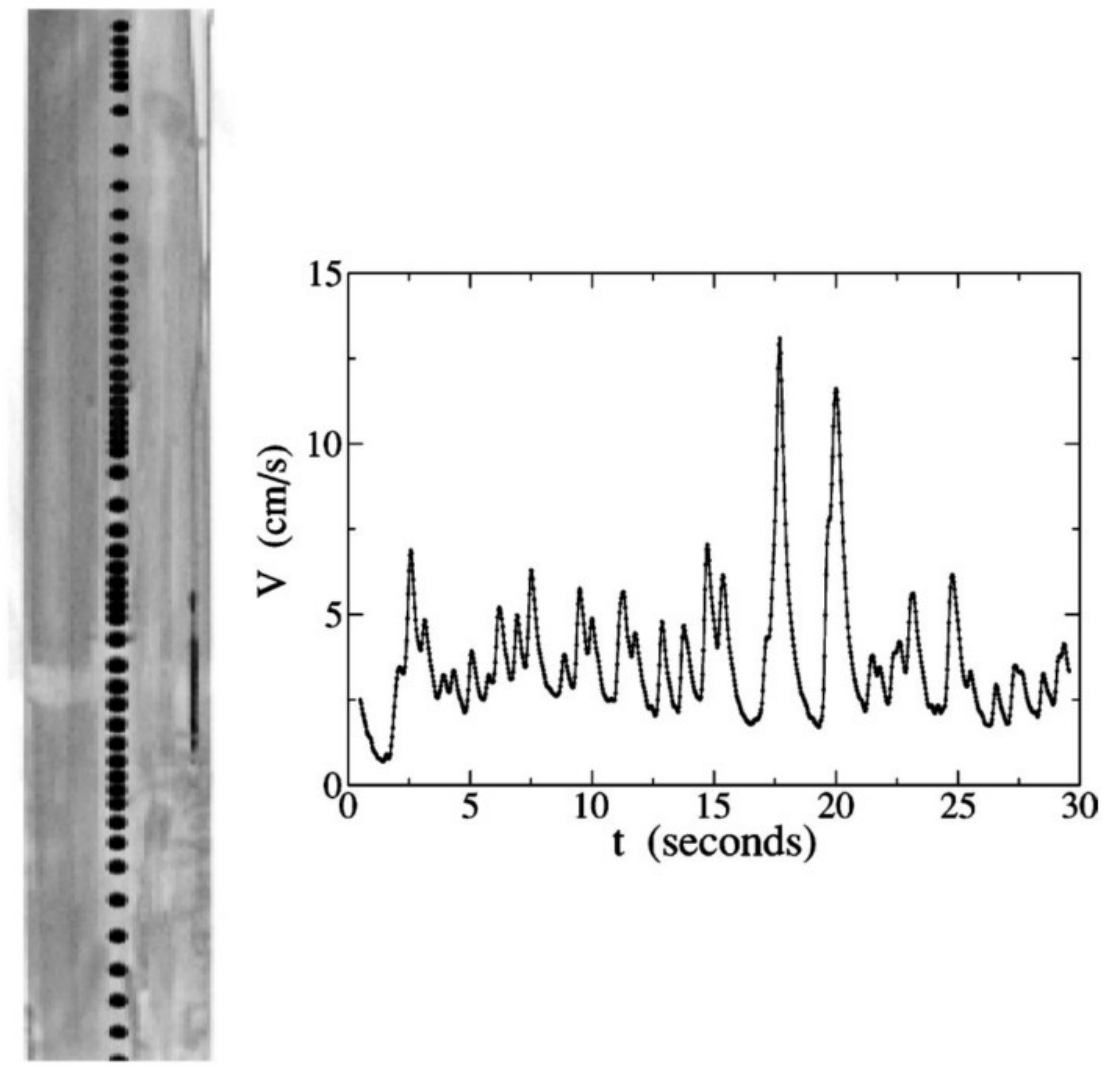}
		\caption{Oscillations of a solid sphere falling through a wormlike micellar fluid}\label{fig:vt}
\end{figure}

However, in applications, it has been found that the falling body behaves with unstable oscillations when it's speed approaching to the terminal velocity \cite{Bird}\cite{Chen}\cite{Chhabra}\cite{Mckinley}\cite{rajagopalan}. Such phenomena can occur in  various cases
such as the high-speed  aircraft,  bullets shot into the water and golf balls falling in shampoo, etc.
This random phenomenon can be observed more clearly when a ball is falling through a non-Newtonian fluid shown in Figure \ref{fig:vt} (e.g. \cite{data}).
\par
The random behaviors of the falling ball depend on the
deformation of the ball and the viscosity of the fluid or air.
You can throw out an iron ball, a tennis or an air-balloon to compare their behaviors. The tennis behaves more randomly  than the iron ball and the air-balloon behaves more randomly than the tennis, even the air is static.

\par
Lee \cite{Lee} takes the fluid as spider nets layer by layer, and add a periodic force caused by the action of the ball which first pushes the spider nets together and  breaks through the spider nets after.
Therefore, the solutions can  be written somewhat like
 $v=v_T (1+d\cos(\omega t))-(v_T-v_0)e^{-\lambda t}$.
A discrete iterative algorithm can also be gotten to simulate the motion of the falling ball.
\par
All the above models are under the framework of Newtonian mechanics and can be described by some second order ODEs. However, people have already found a lot of non-Newtonian fluids such as shampoo, toothpaste, cream, slurry and so on. In fact, all fluids will exhibit it's non-Newtonian properties if the falling ball approaches to the terminal velocity, deforms easily, and the fluid possesses a strong viscosity.
\par
Intuitively, a random term  can be added to the ODE to simulate the random behavior of the falling ball.
The key feature is to find where and what kind of the random terms should be added. It is more important to give a physical interpretation for the random terms.
Our opinion is that the random phenomena are caused by physical relationships themselves. Especially, adding a white noise somewhere is not a good idea.
\par
Wu and Zhang \cite{wuzhang}\cite{wuzhang1} developed a data-driven model by using the Lasso (Least absolute shrinkage and selection operator) to the second order quadric algebraic differential equation (ADE).
Then the model has been compressed and only four terms are remained in the space of $\{v, v^{'}, v^{''}, vv^{'}, vv^{''}, v^{'}v^{''}, v^2, v^{'2}, v^{''2} \}$.
\par
The model is showed  to be
\begin{equation}\label{model-equation}
\beta_9v''^2+\beta_6v''v-\beta_3v''-\beta_1v=0.
\end{equation}
By solving $v$,
\begin{equation}
			\label{model-equation1}
			v=-{\beta_9 v''-\beta_3 \over \beta_6v''-\beta_1}v'',
			\end{equation}
a numerical iterative algorithm
\begin{equation}\label{hysteresis}
			%\label{numerican-algorithm}
			v_{n+1}=-{\beta_9 v''_{n-1}-\beta_3 \over \beta_6v''_{n-1}-\beta_1}v''_{n-1}
			\end{equation}
was got by the time lag. The trajectory of $v$ they simulated by the iterative algorithm (\ref{hysteresis})  seems  to exhibit the chaotic behaviors \cite{wuzhang1}.
These are  some  models  based on the physical laws  to find the important factors which study the physical relations (ADEs) in more and more details and drop the less important factors as errors.
Up to now, the study of the falling ball is a mesoscopic process.
The model (\ref{model-equation}) has already shown the bifurcation behaviors, since the quadric equation has two solutions in every step. Therefore, the trajectory of $v$ actually possesses a deterministic behavior coupled with a random behavior.
Such  trajectories can not be exactly repeatable or reproduced, and can not  be obtained by classical mathematical modeling or Newton's mechanics of ODEs.
\par
What keeps unchanged is the pattern of the motion, but is not the trajectory.
\par
This paper consists of four sections.
In section 2, a data-driven model of the soft Lasso approach is derived under the concept of soft matching of the dynamical equation. After that, a new discrete iterative algorithm to simulate the pattern of a ball falling  in the non-Newtonian fluid is proposed.
In section 3, numerical examples are provided and show very satisfactory results which exhibit the chaotic phenomena, sudden accelerations and the continual random oscillations. The
pattern of the motion is independent of the initial values and is preserved for long time. Some conclusions are given in the last section.

\section{The approach of Data-driven modeling}
In mathematics, we use coordinates to describe the location of a point $(x,y)$ (e.g. Euclidean space $R^2$).
In reality, any point is the mass distribution of a random variable $\xi$. In the following, we call the mass distribution as the point.
In statistics, the expectation $E\xi$ and the variance $\sigma_{\xi}^2$ are used to describe the barycentre ($E\xi$) and size ($\sigma_{\xi}^2$) of a point $\xi$.
Obviously, a point can be described in more detail by its $K$-order moments. Since it is not necessary  to find all moments to determine a point, we will propose a concept of soft matching to define the similarity of two points $\xi$ and $\eta$ at some mesoscopic level.
\begin{definition}\label{soft-matching}
Two points $\xi$ and $\eta$ are called to possess the similarity of order $K$, if $E\xi^k=E\eta^k$ holds for $1\leq k\leq K$. This is denoted as $\xi \simeq_{K} \eta$.
\end{definition}
\begin{proposition}\label{soft-matching1}
If we have already known the distribution of $\eta$ and limited information about $\xi$ (its expectation and variance), from a mesoscopic point of view, a suitable distribution of $\xi$ could be chosen as $E\xi+\sigma_\xi(\eta-E\eta)/\sigma_\eta$. This means we tacitly  take $\xi$ as a affine transformation of the point $\eta$. Since $\xi \simeq_{2} E\xi+\sigma_\xi(\eta-E\eta)/\sigma_\eta$, the soft matching of $\xi$ and $E\xi+\sigma_\xi(\eta-E\eta)/\sigma_\eta$ possesses similarity of order two. We denote it as $\xi\ @\ E\xi+\sigma_\xi(\eta-E\eta)/\sigma_\eta$.
\end{proposition}

Especially, if $\xi$ and $\eta$ obey the same kind of distribution, such as normal distribution or uniform distribution, then they possessing similarity of order $2$ indicates that they are the same point.
\par
Since the concept can be extended to any order of $K$, it can be called a soft matching of two points in certain `mesoscopic’ level. The smaller the $K$, the softer the description of the distribution of $\xi$.
If we know the distribution of $\eta$ and $\xi=c\eta+\varepsilon$, where $\varepsilon$ is a white noise, then roughly speaking, we can tacitly take the distribution of $\xi$  to be
$\xi \ @\ E\xi+{\sqrt{c^2\sigma_\eta^2+\sigma_\varepsilon^2 \over \sigma_\eta^2}}(\eta-E\eta)$.

\par
Take a closer view in the details of the Lasso's result of the model (\ref{model-equation}). The relation of $v$ and $v''$ with the sampling  data is shown in Figure \ref{The data of (v, v'' and the Lasso's result}.
However, the Lasso's result (the hyperbolic curve) could not be taken as a  complete description of the distribution of the data, since the data are not gathering near the curve. So the curve learned by Lasso is only the expectation of the data stream.
\begin{figure}[h]
		\centering		
        %\vspace{-0.4in}
        \includegraphics[width=3in,height=2in]{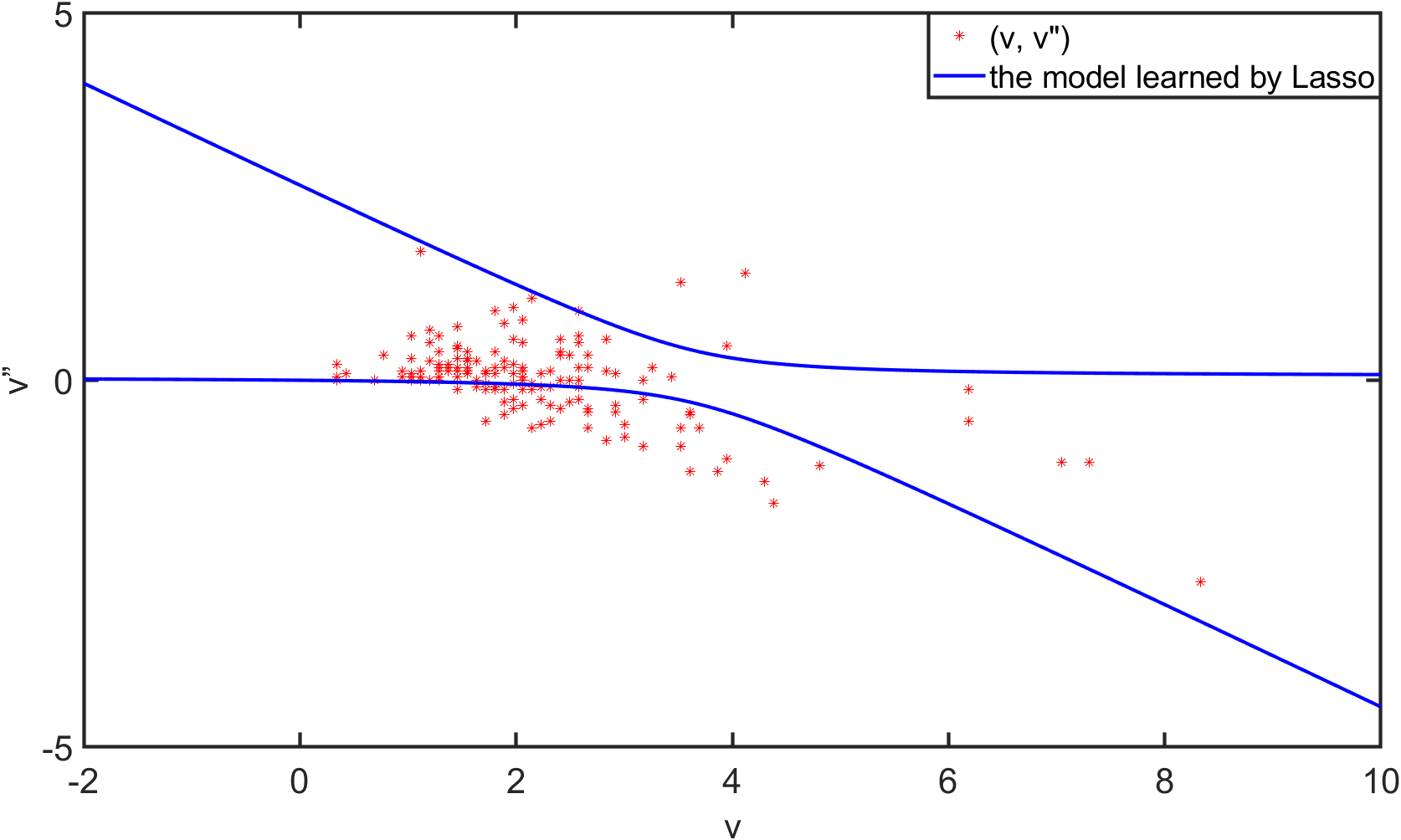}
        %\vspace{-0.7in}
		\caption{The data of $(v,v'')$ and the Lasso's result \label{The data of (v, v'' and the Lasso's result}}	
\end{figure}
\par
Since the trajectory of the velocity of a ball falling in the  non-Newtonian fluid can not be reproduced, we hope to reproduce the pattern only.
The problem is, how to define the similarity or equivalency  of the  patterns used in the pattern classification of the trajectory of the velocity in Figure \ref{fig:vt}.
\par
However, pattern classification of random  trajectories is still a mesoscopic problem. What we can reproduce is a similar trajectory at some mesoscopic level described in the Proposition \ref{soft-matching1} of soft matching.
The model (\ref{model-equation}) learned by Lasso is reformed
by dividing the $\beta_9\omega^4$ and denoted as $\zeta$
\begin{equation}
\label{new-model1}
\begin{array}{rl}
\zeta:= &({v'' \over \omega^2})^2+2{v''\over  \omega^2}v -  2c{v''\over \omega^2}-2c(1-d){v}\\
=&({v''\over \omega^2}+(v-c))^2-c^2 - (v-dc)^2+(dc)^2,\\
	\end{array}		\end{equation}
 where $\omega^2=\beta_6/2\beta_9$, $c=\beta_3/\beta_6$ and  $1-d=2\beta_1\beta_9/\beta_3\beta_6$.
What has been learned by  Lasso in \cite{wuzhang} is to find the coefficients  to minimize $E\zeta^2$, and takes the result as $\zeta=0$. However, since the vectors $\{v''^2,v''v, v'', v\}$ of data are linear independent, $\zeta$ is not a pure mathematical zero but is still a random distribution. Therefore, we can only get a solution $\zeta$ such that $E\zeta=0$, and we will call this approach as soft Lasso. Denote the expectation of the velocity $v$ as $\bar{v}$, and let $\xi=({v''\over \omega^2}+(v-\bar{v}))$ and $\eta= (v-\bar{v})$. Then the formula (\ref{new-model1}) can be written as
\begin{equation}\begin{array}{rl}
0=E\zeta=
&E({v''\over \omega^2}+(v-\bar{v}))^2 - E(v-\bar{v})^2-2\bar{v}c(1-d)\\
&=E\xi^2-E\eta^2-2\bar{v}c(1-d),\\
\end{array}	
\end{equation}
since $E\xi=0$ and $E\eta=0$. Therefore, we can take the point $\xi$ as $\eta+\varepsilon$, where  $\varepsilon$ is a white noise with $E\varepsilon=0$  and $E\varepsilon^2=2\bar{v}c(1-d)$.
\par
Base on the Proposition \ref{soft-matching1} of soft matching in the mesoscopic point of view, we have
$$\xi\  @\   \sqrt{\sigma^2_\eta+2\bar{v}c(1-d)    \over \sigma^2_\eta}     \eta,$$
recalling $\xi=({v''\over \omega^2}+(v-\bar{v}))$,
\begin{equation}\label{newmainresult}
({v''\over \omega^2}+v-\bar{v})\  @\   \sqrt{\sigma^2_\eta+ 2\bar{v}c(1-d)    \over \sigma^2_\eta}\eta.
\end{equation}
This leads to a discrete algorithm for soft Lasso
\begin{equation}\label{discrete-soft-Lasso}
  v_{n+1}-2v_n+v_{n-1}+2\omega^2(v_n-\bar{v})\
   @\ 2\omega^2\sqrt{\sigma_\eta^2+2\bar{v}c(1-d) \over \sigma_\eta^2}\eta.
\end{equation}
Based on the ergodic theory ($\bar{v}$ approaches to the terminal velocity $v_T$), we take the discrete Stokes's equation
\begin{equation}\label{discrete-newton-stokes}
   v_{n+1}-v_{n-1}+2\lambda(v_n-\bar{v})=0
\end{equation}
into account to simulate the macroscopic Stokes equation $v^{'}=\lambda (\bar{v}-v)$. This derives $\lambda=\sqrt{\frac{E(v')^2}{E(v-\bar{v})^2}}$ from the data of velocity $v$.
\par
Since the trajectory of $v$ comes from a coupled dynamic system with deterministic and random terms,
then  combining the soft Lasso term (\ref{discrete-soft-Lasso}) and the Stokes's term (\ref{discrete-newton-stokes}), we
derive a soft dynamical discrete iterative algorithm for a ball falling in a non-Newtonian fluid under the concept of soft matching,
\begin{equation}\label{newmain}
\begin{array}{l}
(1+\epsilon)v_{n+1}-2(1-\omega^2-\epsilon\lambda)v_n+(1-\epsilon)v_{n-1}
-2(\omega^2+\epsilon\lambda)\bar{v}\\
   @\ 2\omega^2(\sqrt{\sigma_\eta^2+2\bar{v}(1-d)c \over \sigma_\eta^2})\eta.
   \end{array}
\end{equation}
That means the weights of the soft Lasso term and the Stokes’s term are $1/(1+\epsilon)$ and $\epsilon/(1+\epsilon)$ respectively.
The parameters $\bar{v}, \omega, c, d, \sigma_{\eta}, \lambda$ can be gotten by statistics and Lasso approach from the data of $v$.
We will solve the equation (\ref{epsilonequation}) to find the parameter $\epsilon$.
The formula (\ref{newmain}) can be reformed to be
\begin{equation}\label{newmain1}
(1+\epsilon)(v_{n+1}-\bar{v})\ @ \ 2(1-\omega^2-\epsilon\lambda)(v_n-\bar{v})-(1-\epsilon)(v_{n-1}-\bar{v})
   +2\omega^2(\sqrt{\sigma_\eta^2+2\bar{v}(1-d)c \over \sigma_\eta^2})\eta.
\end{equation}

Taking the expectations of both sides of (\ref{newmain1}), we find formula (\ref{newmain1}) is unbiased.
Taking the covariance of  $v_n-\bar{v}$ and formula (\ref{newmain1}), we get
\begin{equation}
(1+\epsilon)\gamma(1)=2(1-\omega^2-\epsilon\lambda)\gamma(0)-(1-\epsilon)\gamma(1),
\end{equation}
where $\gamma(0)=E(v_n-\bar{v})^2=var(v_n)$ and $\gamma(1)=E(v_n-\bar{v})(v_{n+1}-\bar{v})=E(v_n-\bar{v})(v_{n-1}-\bar{v})$.
More discussions of the
covariance can be found in \cite{Yaglom}.
\par
Therefore,
\begin{equation}\label{relation0and1}
\gamma(1)=(1-\omega^2-\epsilon\lambda)\gamma(0).
\end{equation}
Simplifying the variances of both sides of (\ref{newmain1}) and using (\ref{relation0and1}), we get
\begin{equation}
\begin{array}{l}
\epsilon\gamma(0)=(1-\omega^2-\epsilon\lambda)^2\epsilon\gamma(0)
  + \omega^4{\sigma_\eta^2+2\bar{v}(1-d)c \over \sigma_\eta^2}var(\eta).
   \end{array}
\end{equation}
Then the equation of $\epsilon$
\begin{equation}\label{epsilonequation}
(2-\omega^2-\epsilon\lambda)(\omega^2+\epsilon\lambda)\epsilon=\omega^4{\sigma_\eta^2+2\bar{v}(1-d)c \over \sigma_\eta^2}
\end{equation}
holds in order to reproduce the variance of the distribution of velocity $v$ which means $\gamma(0)=var(\eta)$.
Now we just need to get an empirical distribution of $v$ to simulate the behavior of a falling ball by our main algorithm
(\ref{newmain}). In our problem, the data of velocity are collected from the physical experiment in \cite{data} which are showed in Figure \ref{fig:vt}.

\section{Numerical Examples}
\begin{figure}[h]
		\centering
		\includegraphics[width=3in,height=2in]{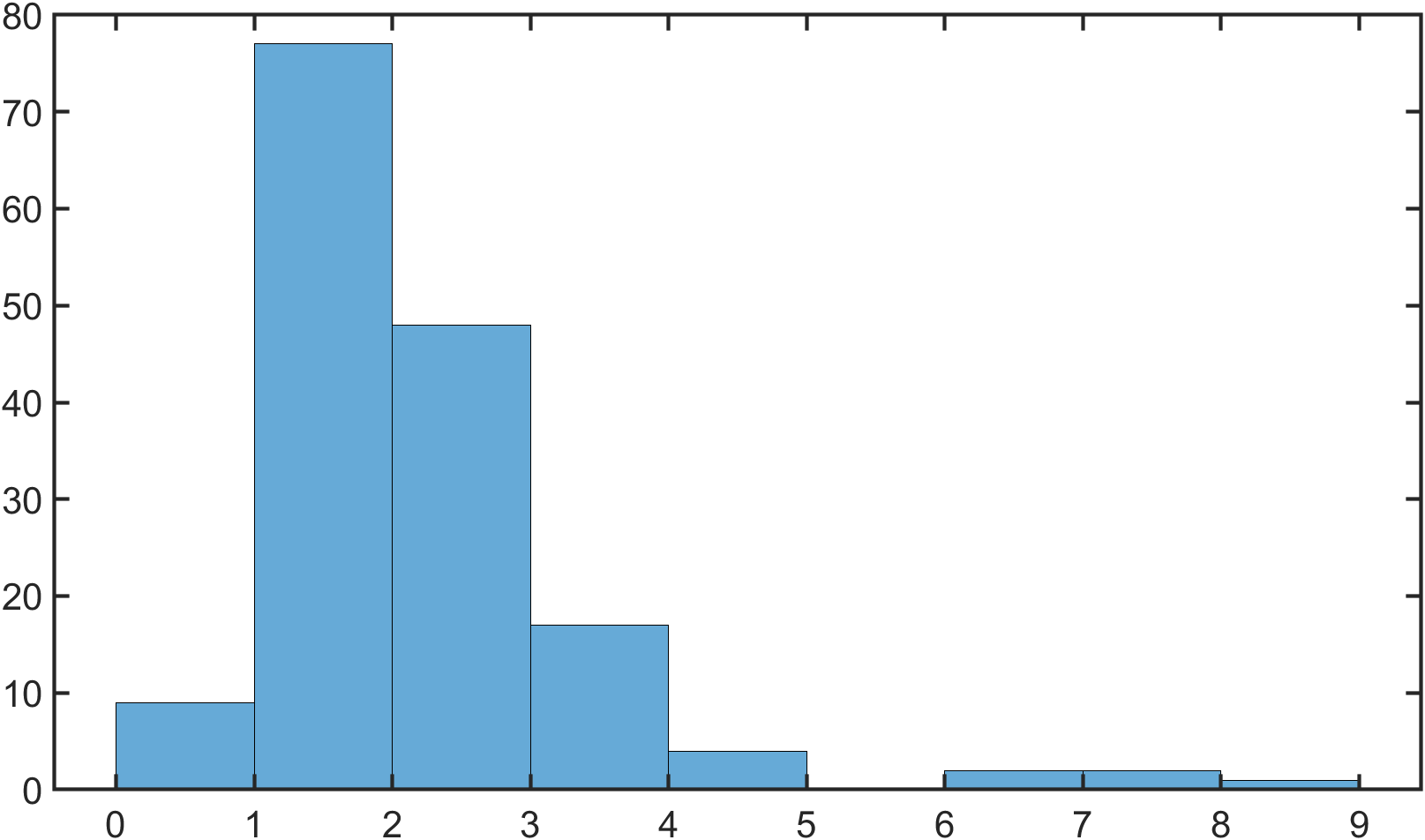}
		\caption{The histogram of velocity samples \label{v-histogram}}	
\end{figure}

Based on the histogram of $v$ shown in Figure \ref{v-histogram},
the superposition of two uniform, two Wigner semicircle distributions and two triangular distributions on the interval $[1, 4] \cup [7,9]$ are used first as a rule of thumb to describe the distribution of the  velocity $v$ in Example 1. Then considering the expectation and variance of the velocity samples, we give more precise distributions of the velocity $v$ in Example 2. The experiments show very satisfactory results, which exhibit the chaotic phenomena, the sudden acceleration and the continual random oscillations.

\subsection{CDFs of the three distributions}
We first introduce some cumulative distribution functions (CDFs).
The  uniform distribution  on $(a,b)$ centered at $(a+b)/2$ is
$$F_u(v, (a, b))={1\over 2}+{1\over 2}({2v-a-b \over b-a})=\frac{v-a}{b-a},\ \ \ \ v\in (a,b),$$
with the variance of $(b-a)^2/12$.
\par
The Wigner semicircle distribution on $(a,b)$ centered at $(a+b)/2$ is
$$F_W(v, (a, b))=\frac{1}{2}+{1\over \pi}\arcsin({2v-a-b \over b-a})+{1\over \pi}({2v-a-b \over b-a})\sqrt{1-({2v-a-b \over b-a})^2}, \ \ \ v\in (a,b),$$
with the variance of $(b-a)^2/16$.
\par
The triangular distribution on $(a,b)$ with mode (peak location) $c$ ($a\leq c \leq b$) is
\begin{equation}
F_t(v,(a, b, c))=
\left\{
\begin{array}{lr}
\frac{(v-a)^2}{(b-a)(c-a)}, & a<v\leq c, \\
1-\frac{(b-v)^2}{(b-a)(b-c)}, & c<v\leq b.\\
\end{array}
\right.
\end{equation}
Its expectation is $(a+b+c)/3$ and variance is $(a^2+b^2+c^2-ab-ac-bc)/18$.
The probability density functions (PDFs) of the three distributions are all supported on the interval $[a, b]$. For more details about the distributions,
we refer to references \cite{beyond}\cite{probability}.
\par
We have used the above three distributions and even the $\beta-$distribution as the empirical distribution of $v$ to simulate the behavior of a falling ball, and get some results which have shown the random  oscillations. However, in order to  reproduce the physical phenomena of sudden accelerations which appear in Figure \ref{fig:vt}, we adopt the superposition of two such distributions.

\subsection{Example 1: Distributions as a rule of thumb}
\begin{figure}[H]
\centering
\subfigure[]{
\includegraphics[width=5cm]{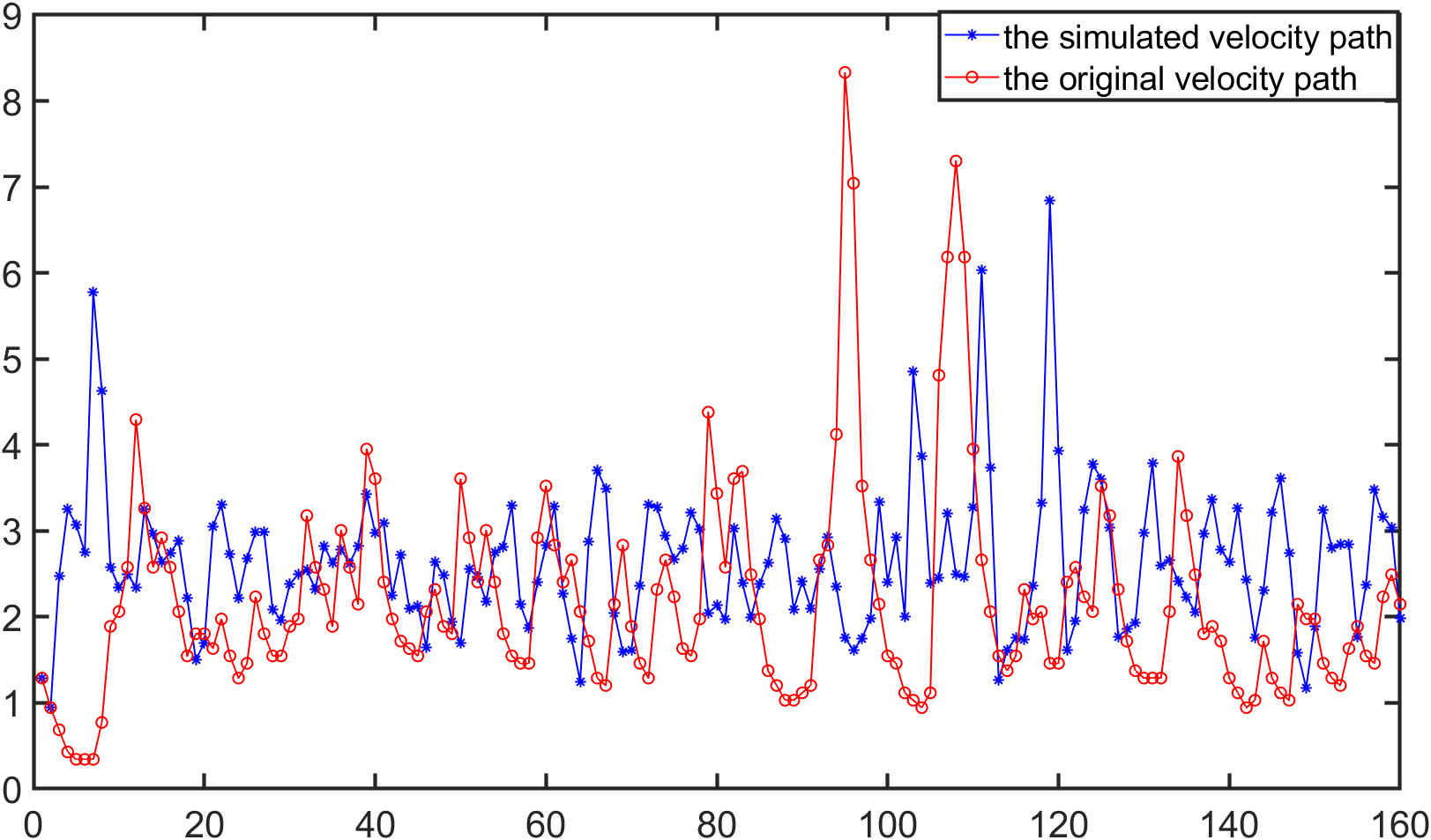}
}
\quad
\subfigure[]{
\includegraphics[width=5cm]{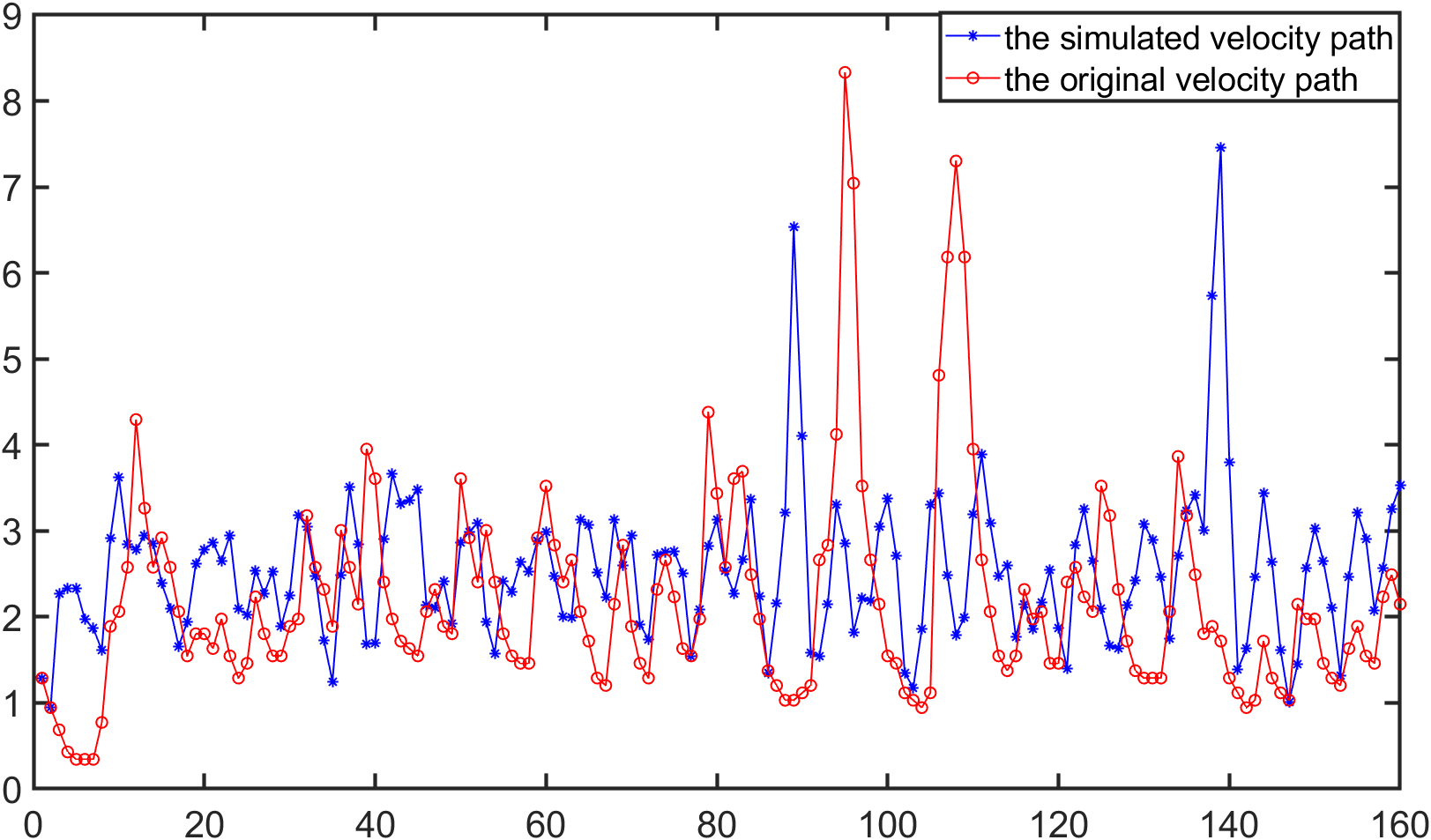}
}
\quad
\subfigure[]{
\includegraphics[width=5cm]{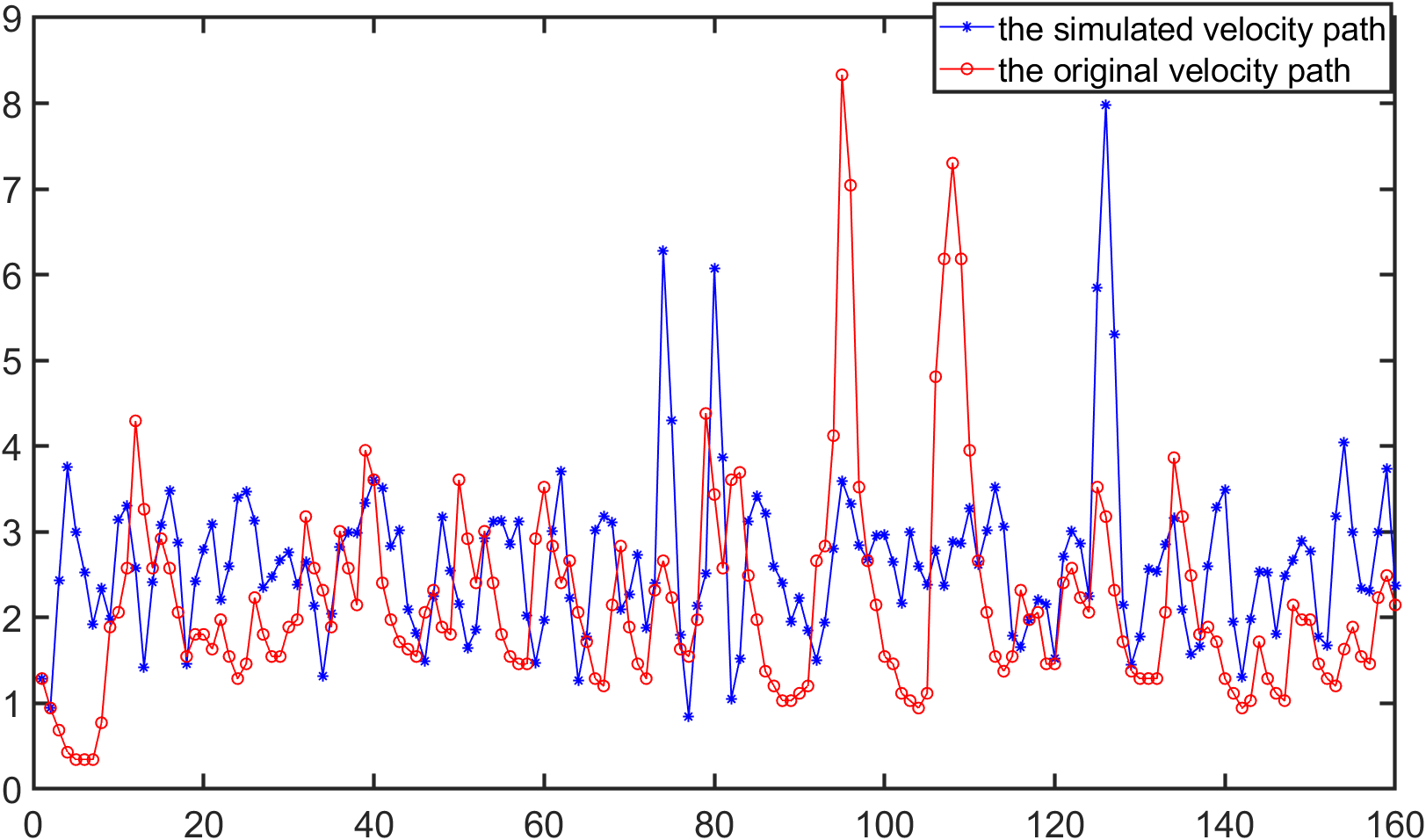}
}
\quad
\subfigure[]{
\includegraphics[width=5cm]{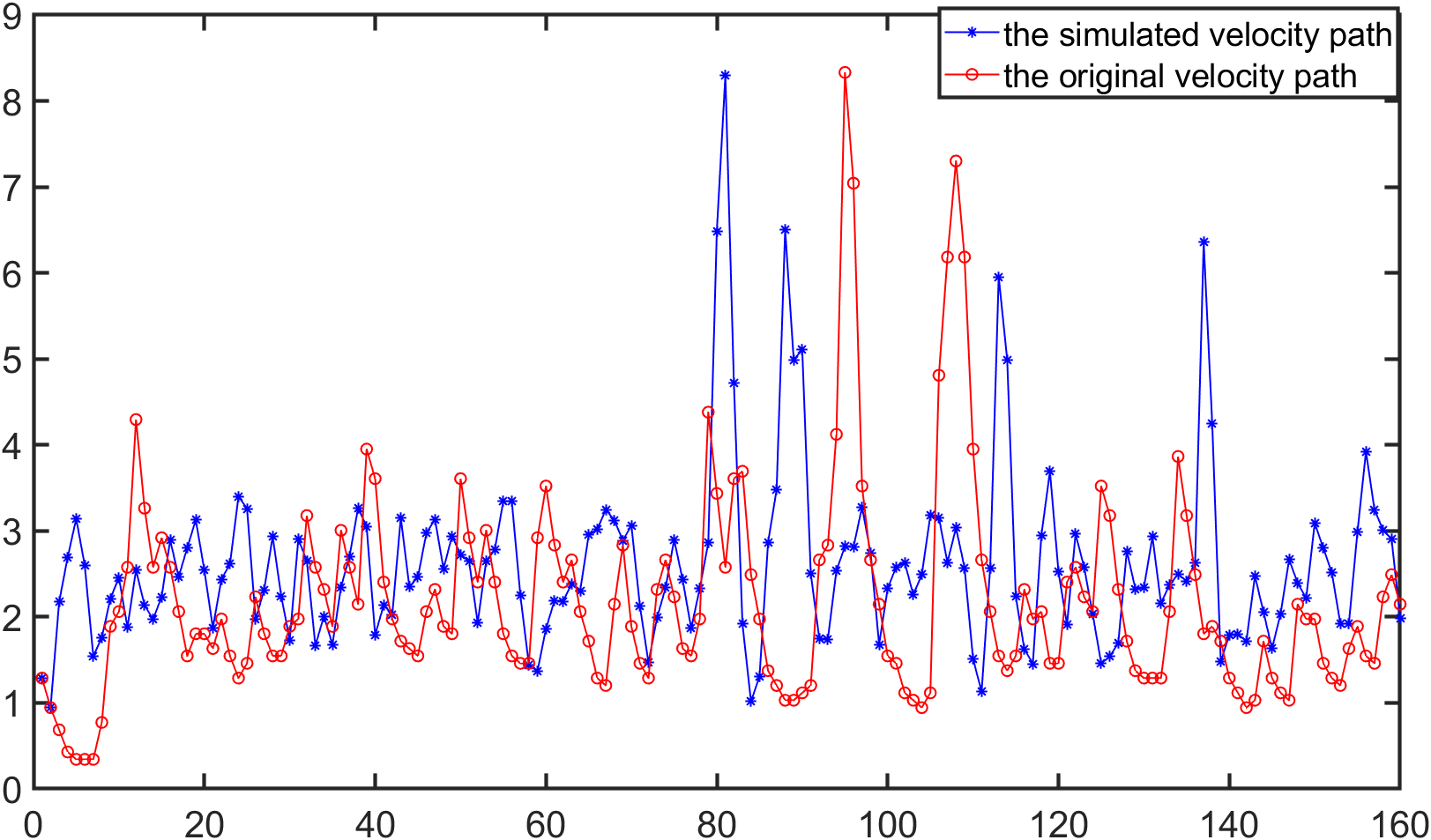}
}
\caption{The simulated velocity under the distribution $F_{W2}(v, 0.9683, (1, 4), (7, 9))$.}\label{twosc1}
\end{figure}

\begin{figure}[H]
\centering
\subfigure[]{
\includegraphics[width=5cm]{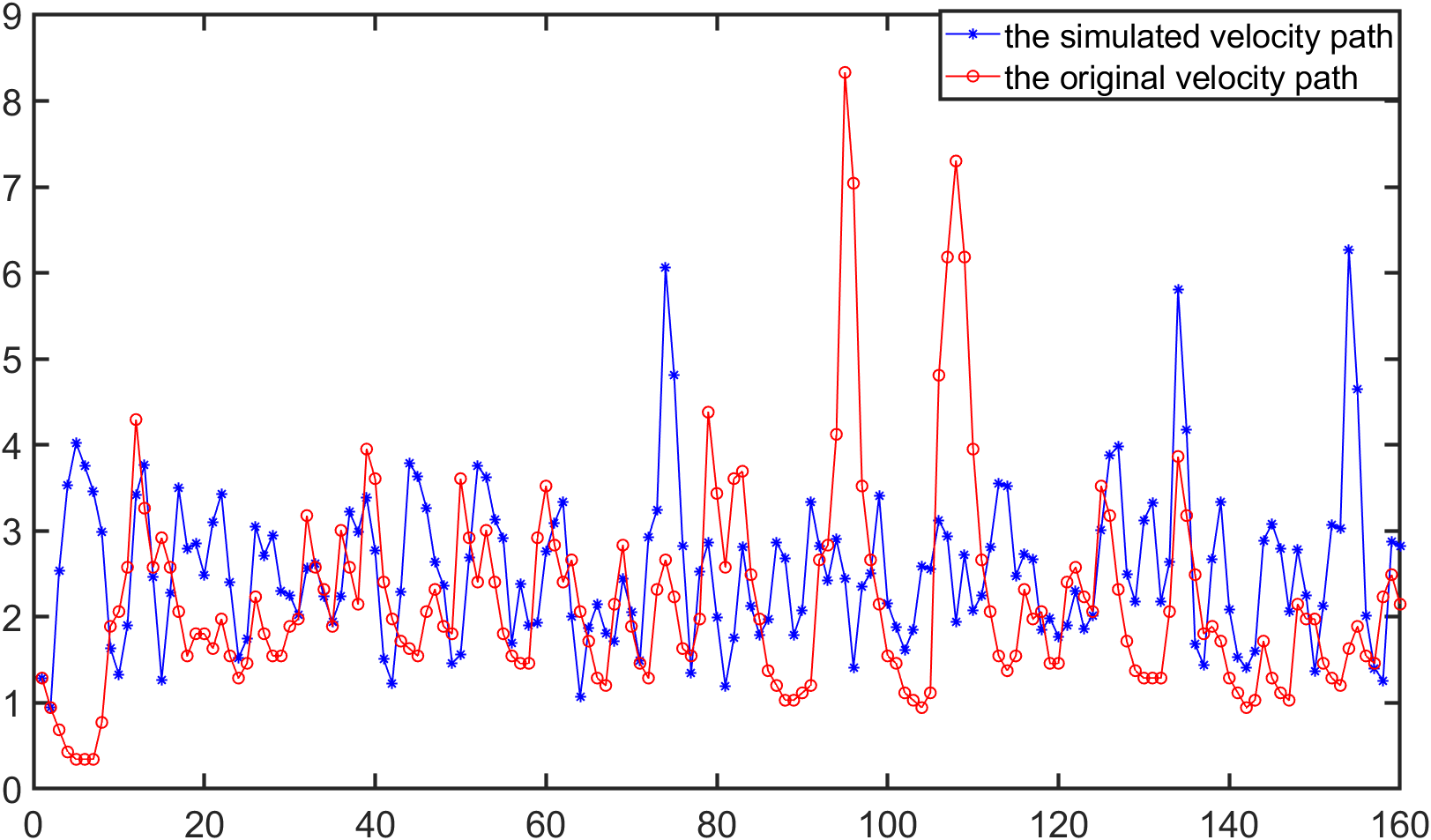}
}
\quad
\subfigure[]{
\includegraphics[width=5cm]{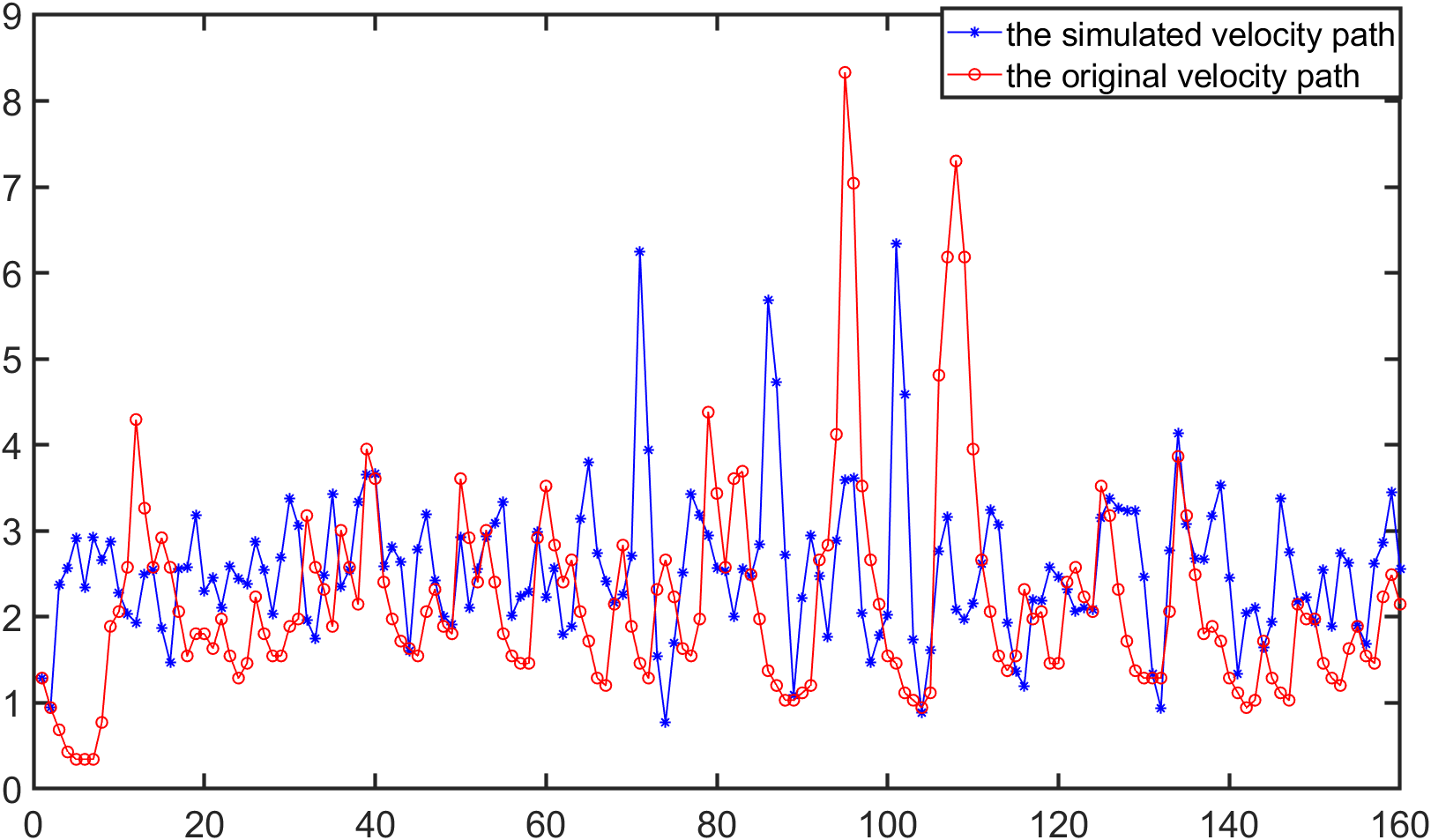}
}
\quad
\subfigure[]{
\includegraphics[width=5cm]{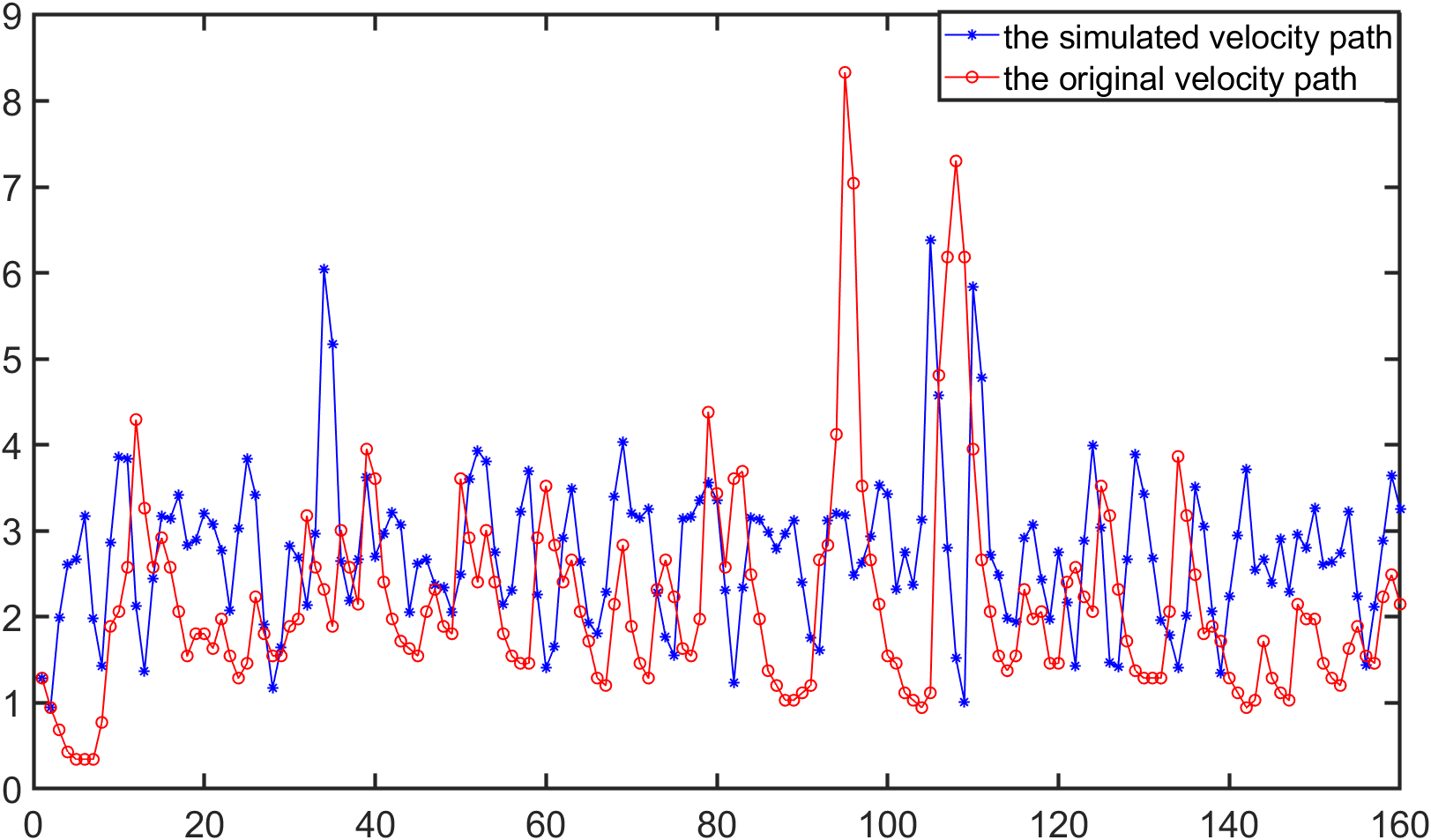}
}
\quad
\subfigure[]{
\includegraphics[width=5cm]{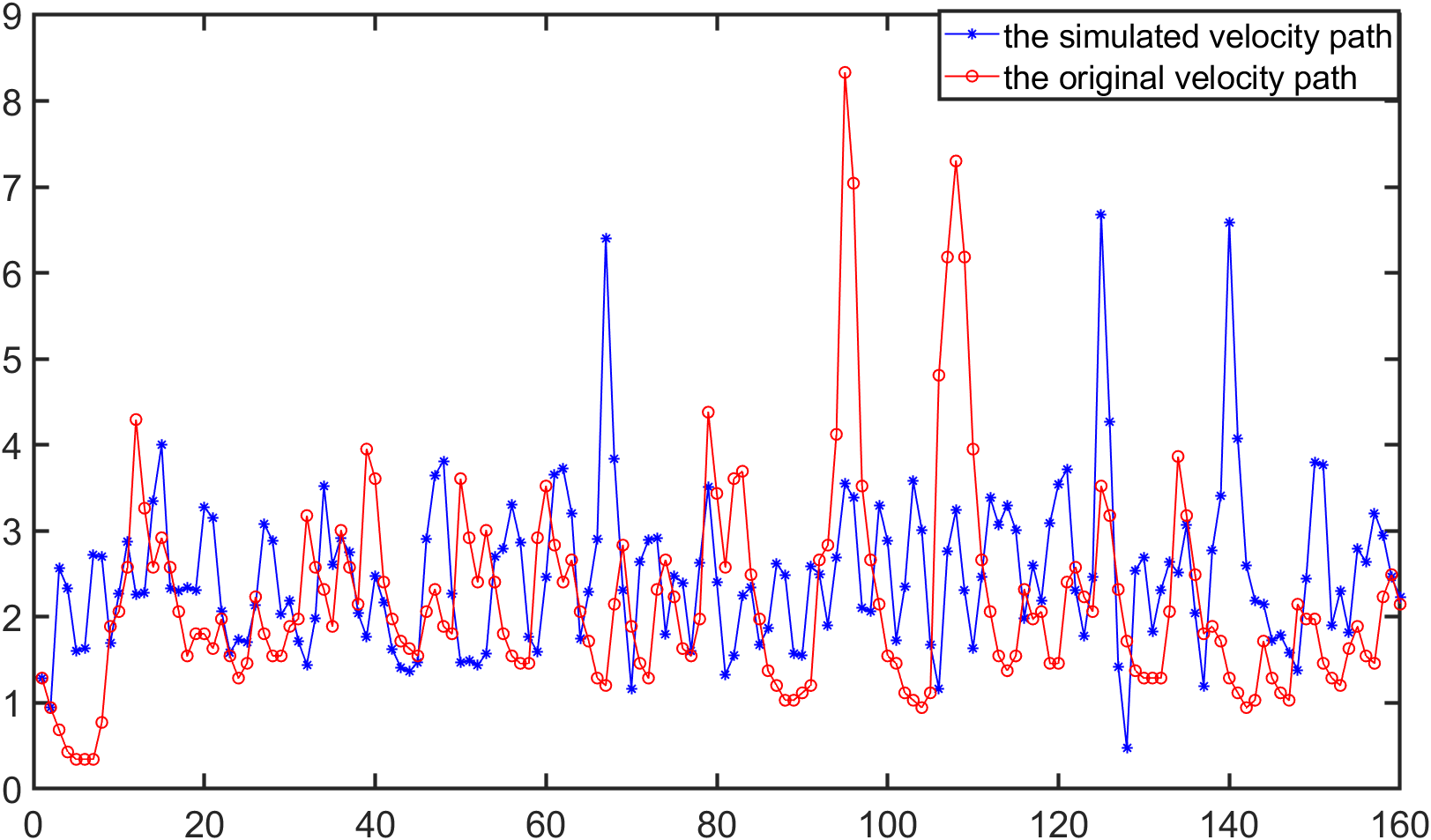}
}
\caption{The simulated velocity under the distribution $F_{u2}(v, 0.9683, (1, 4), (7, 9))$.}\label{twoun1}
\end{figure}

\begin{figure}[H]
\centering
\subfigure[]{
\includegraphics[width=5cm]{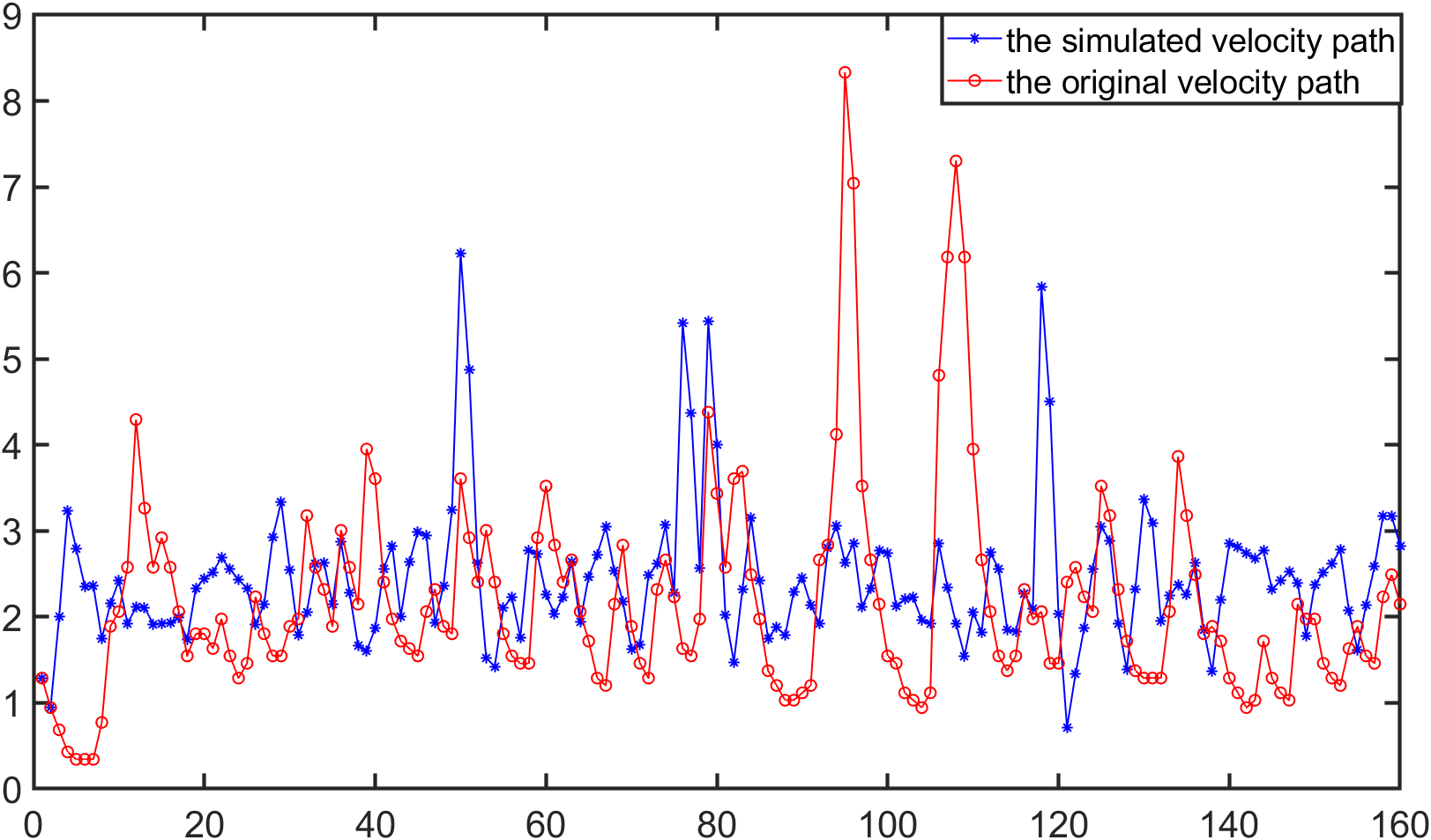}
}
\quad
\subfigure[]{
\includegraphics[width=5cm]{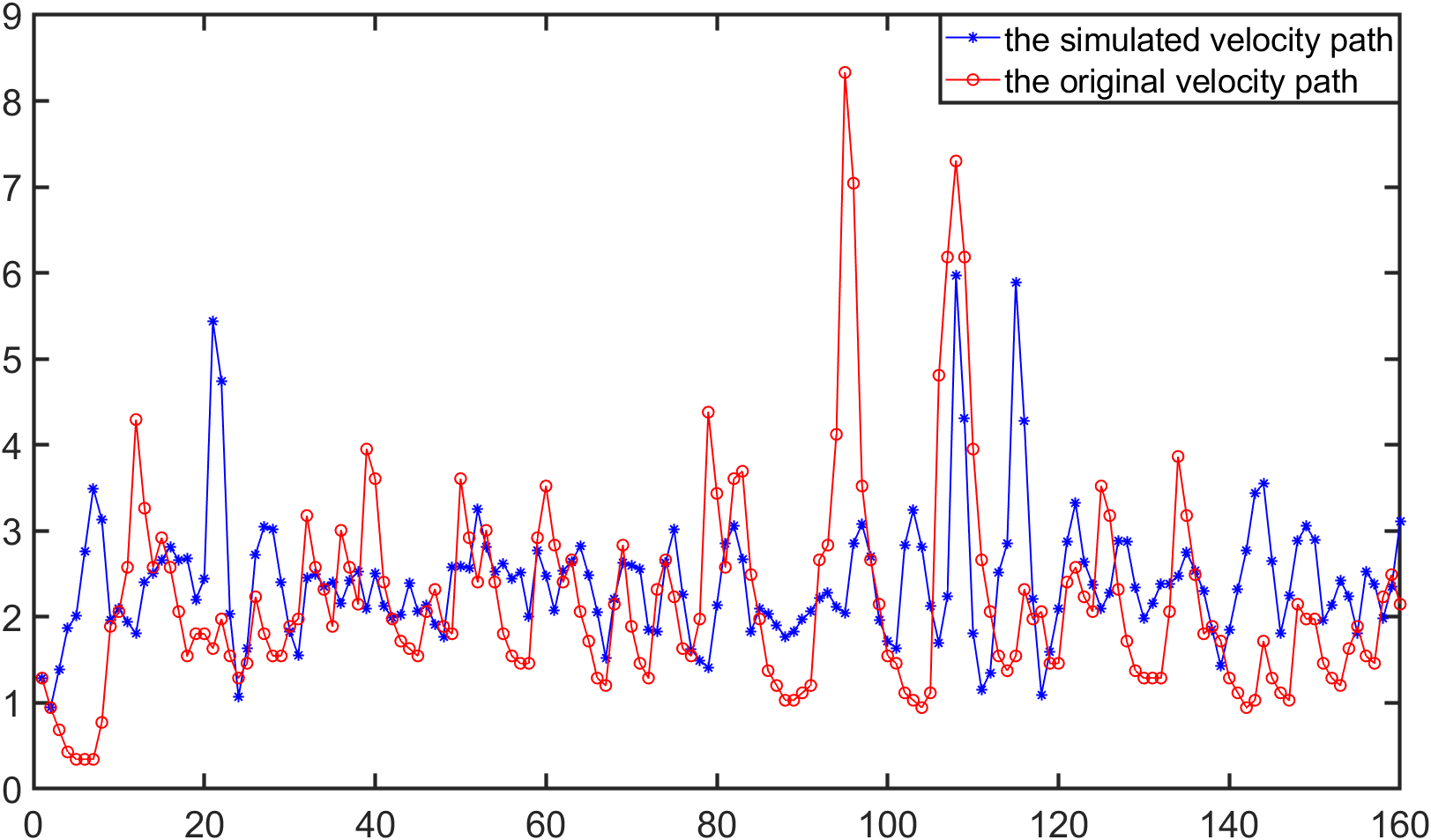}
}
\quad
\subfigure[]{
\includegraphics[width=5cm]{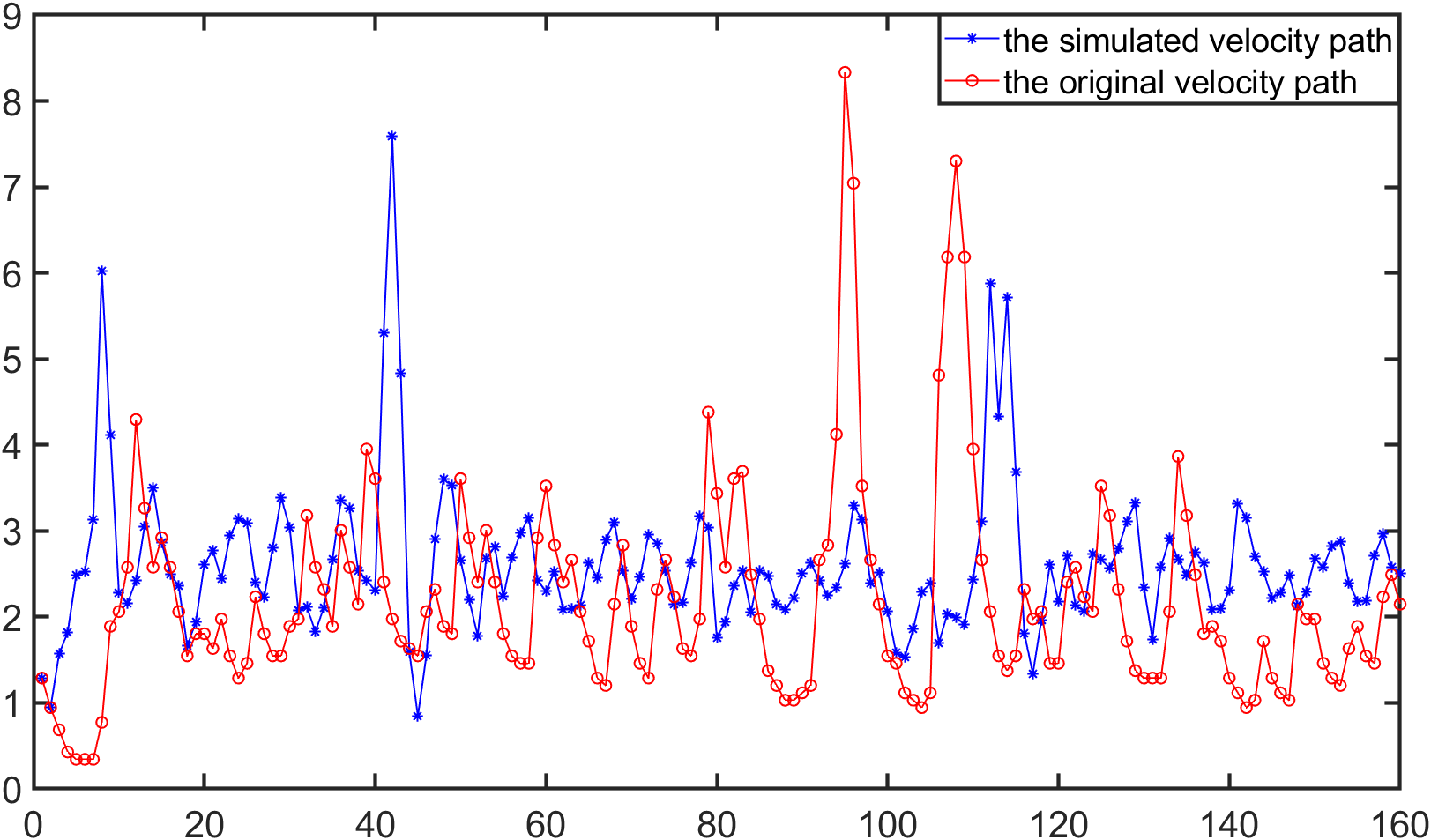}
}
\quad
\subfigure[]{
\includegraphics[width=5cm]{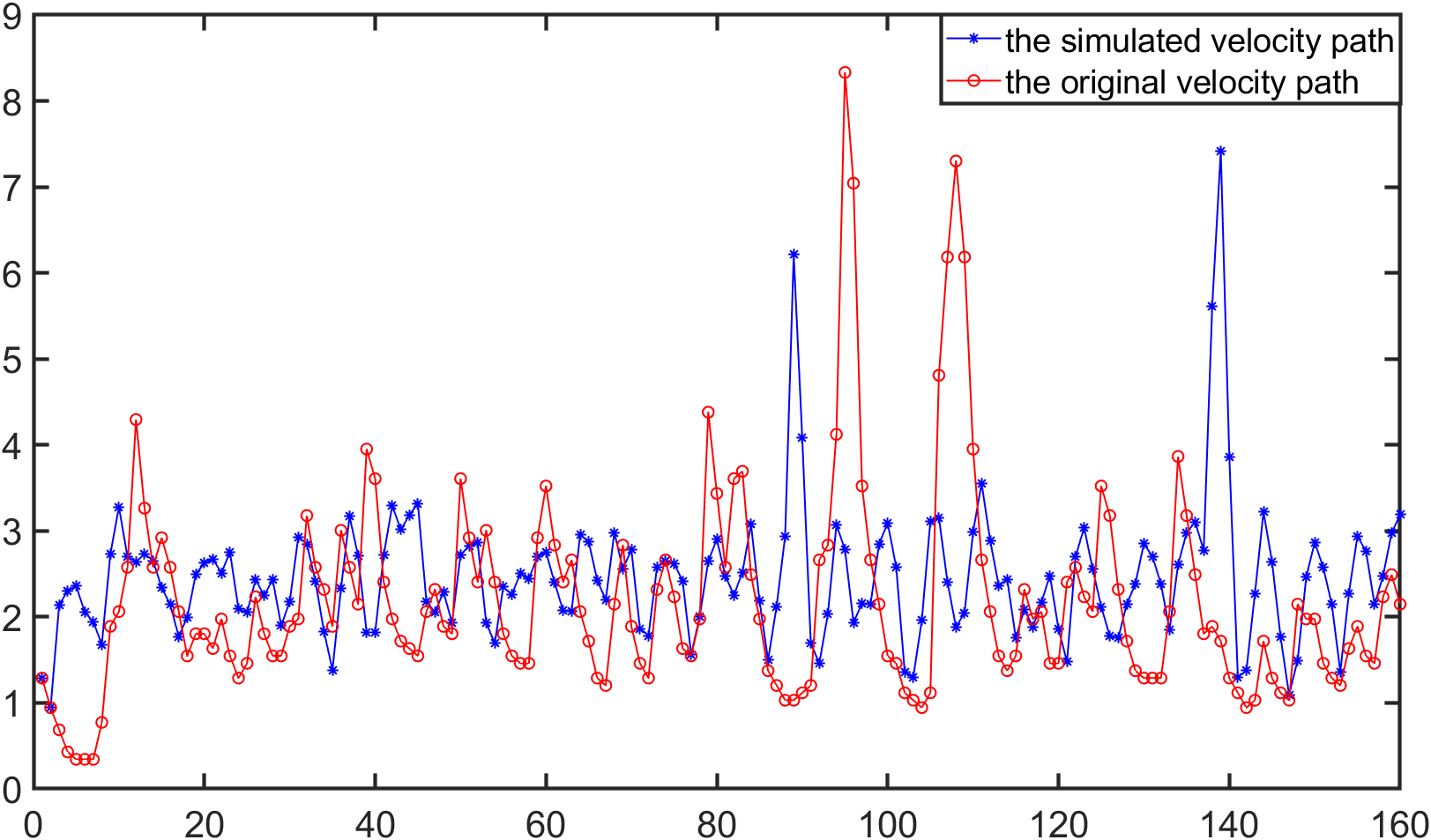}
}
\caption{The simulated velocity under the distribution $F_{t2}(v, 0.9683, (1, 2.5, 4), (7, 8, 9))$.}\label{twotr1}
\end{figure}

From Figure \ref{fig:vt}, we observe that
the samples of velocity mainly distribute in  two parts, which would be regarded as the intervals $[1, 4]$ and $[7, 9]$ as a rule of thumb when getting into the equilibrium state.
So the empirical distribution of $v$ could be chosen as the superposition of two uniform, two Wigner semicircle or two triangular distributions ($F_{u2}, F_{W2}, F_{t2}$) supported on the intervals $[1, 4]$ and  $[7,9]$ with the weights determined by the ratio of the sudden acceleration in equilibrium state. Then
using the main algorithm (\ref{newmain}), the following results are obtained and showed in Figures \ref{twosc1}, \ref{twoun1}, \ref{twotr1} by adopting the optimized parameter of $\epsilon$ (and throughout the paper).

Numerical results above demonstrate the pattern of a falling ball, such as the chaotic phenomena, sudden accelerations and continual random oscillations. It seems that the superposition of two uniform distributions with the largest  variances of the three  distributions seems to be better.  However,
the three empirical distributions have not achieved to the aim of reproducing  the expectation and variance of the samples which will be discussed in next subsection.
\subsection{Example 2: Reproducing  expectation and variance}
\begin{figure}[H]
\centering
\subfigure[]{
\includegraphics[width=5cm]{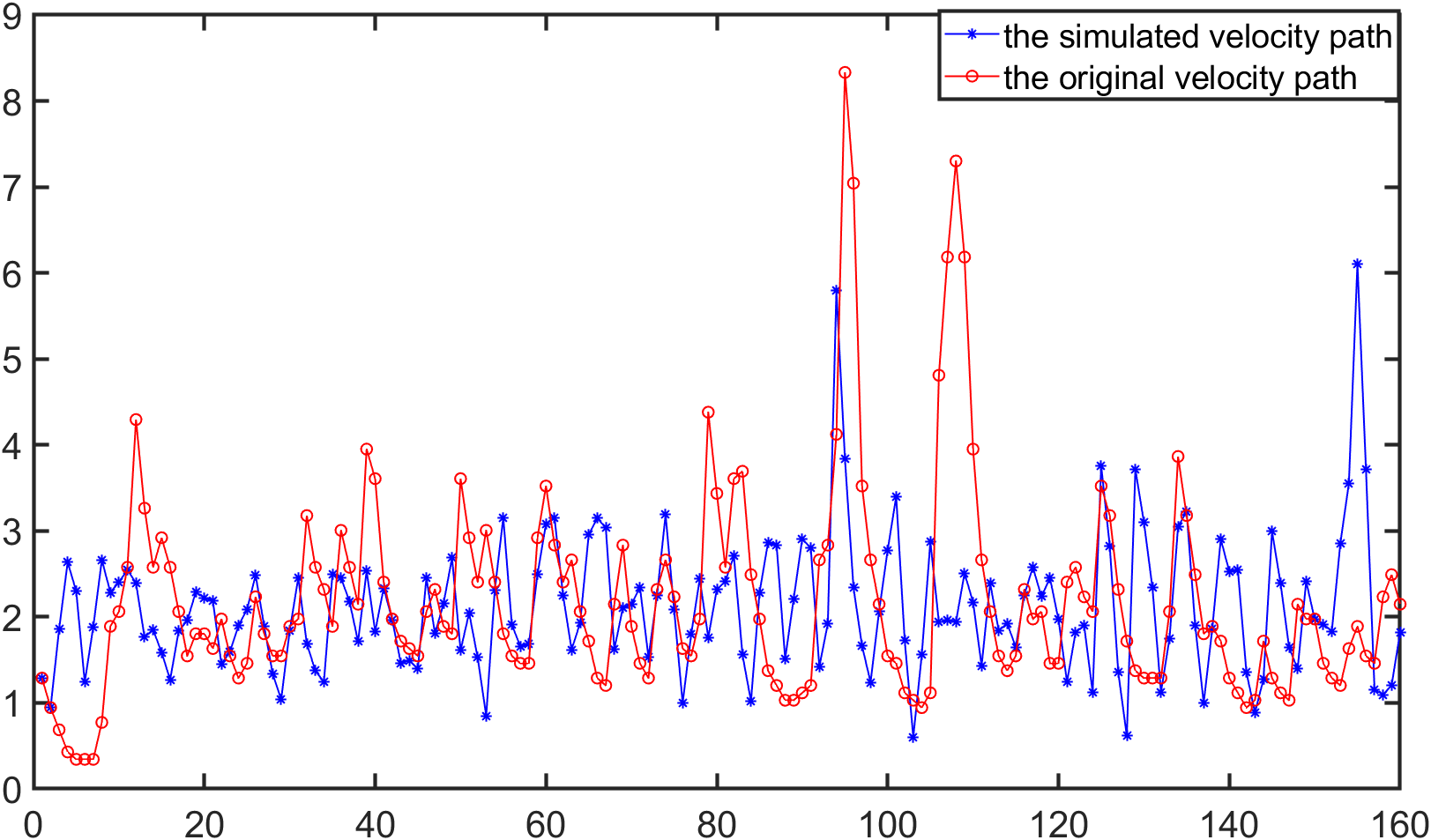}
}
\quad
\subfigure[]{
\includegraphics[width=5cm]{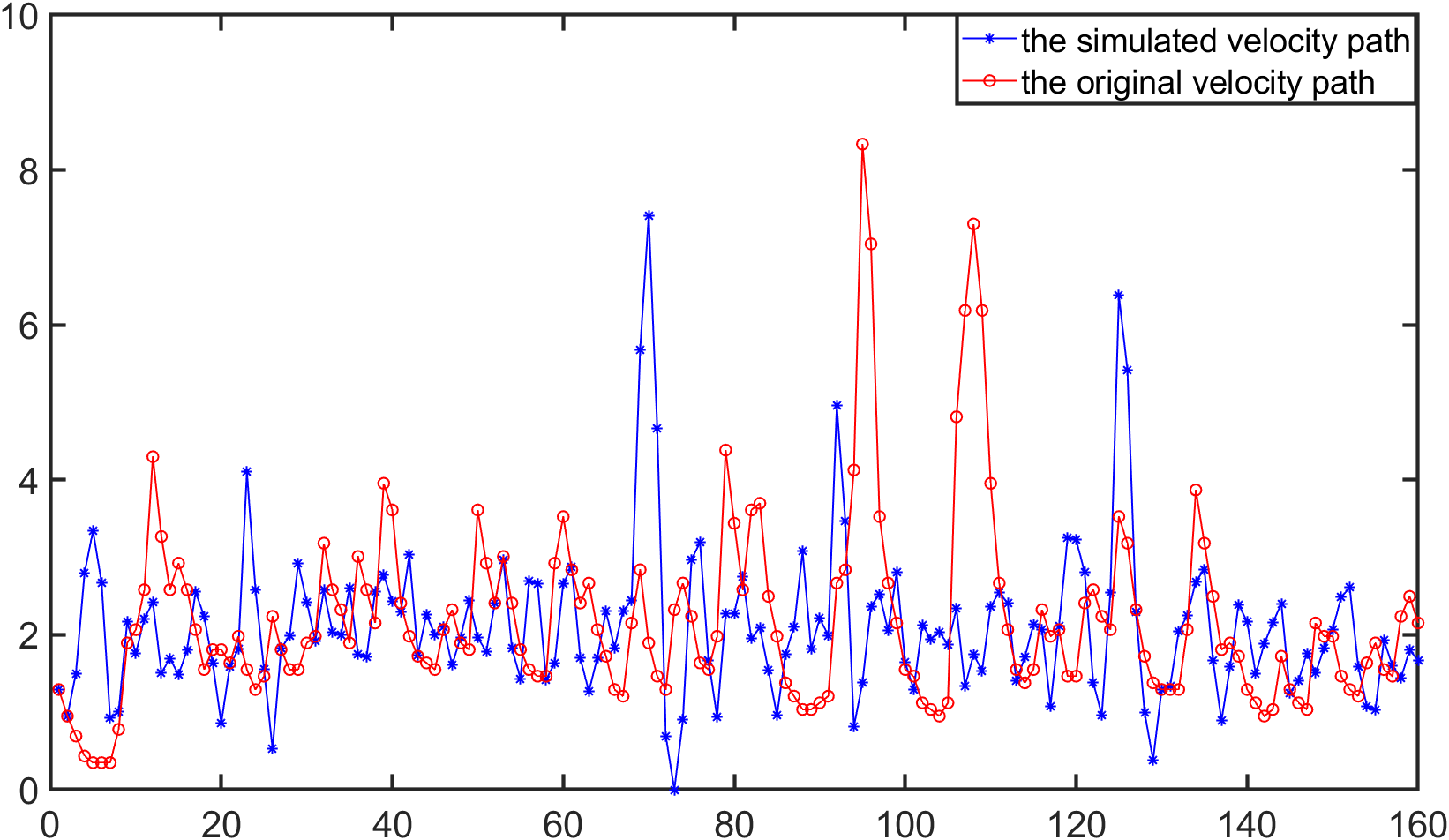}
}
\quad
\subfigure[]{
\includegraphics[width=5cm]{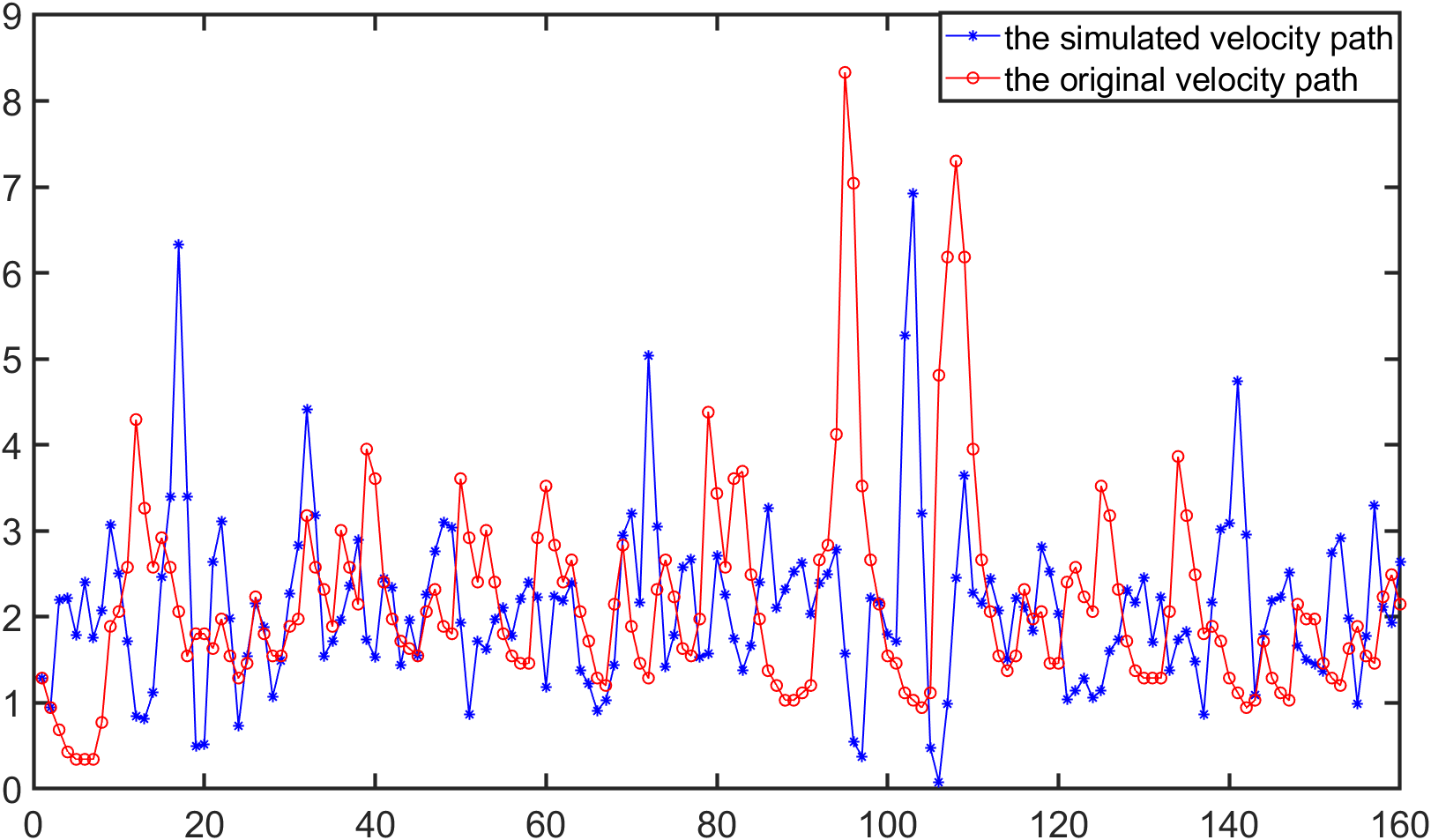}
}
\quad
\subfigure[]{
\includegraphics[width=5cm]{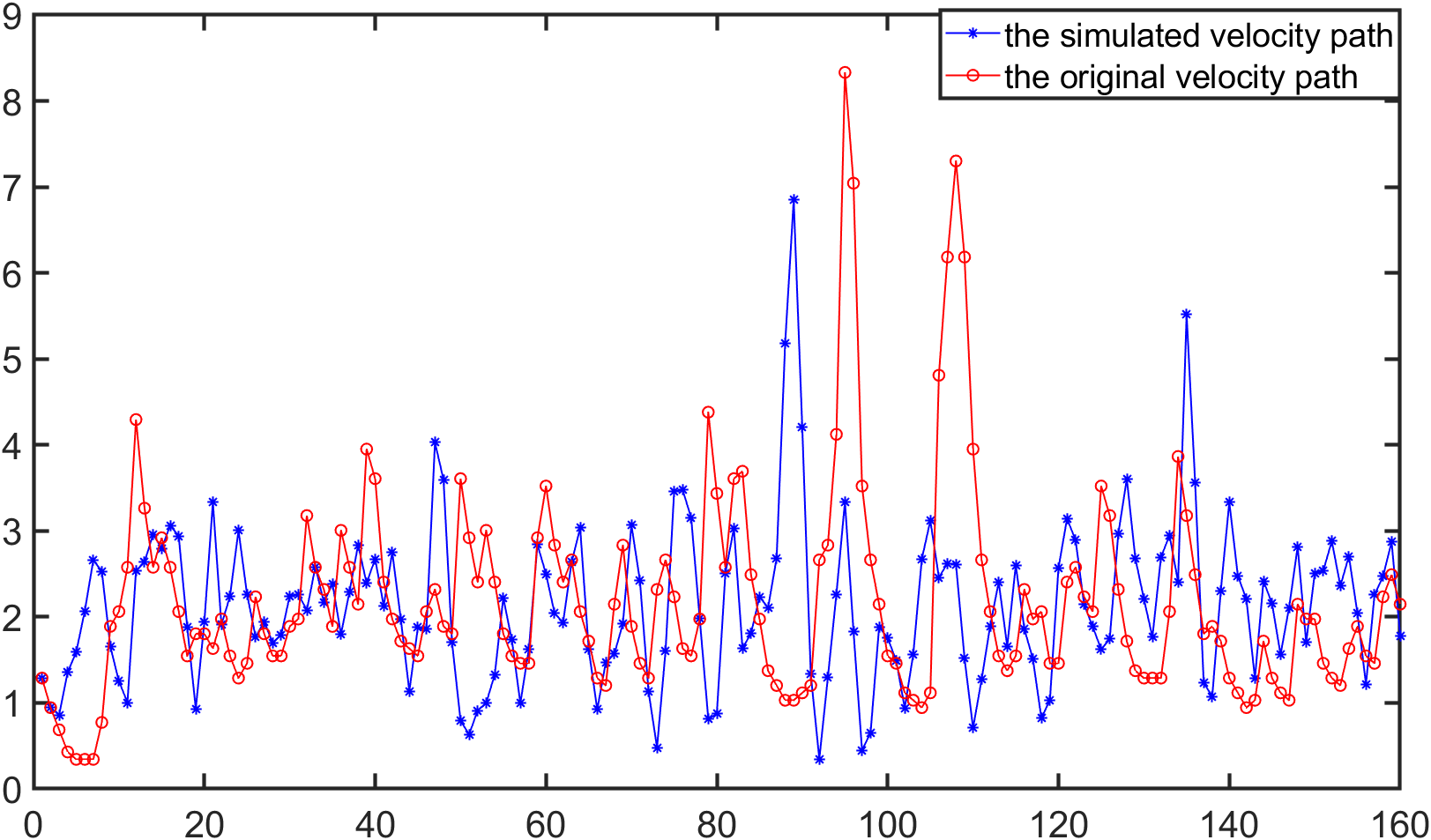}
}
\caption{The simulated velocity under the distribution $F_{W2}(v, 0.9683, (0.3155, 3.7706), (5.2216, 8.7973))$.}\label{twosc2}
\end{figure}

\begin{figure}[H]
\centering
\subfigure[]{
\includegraphics[width=5cm]{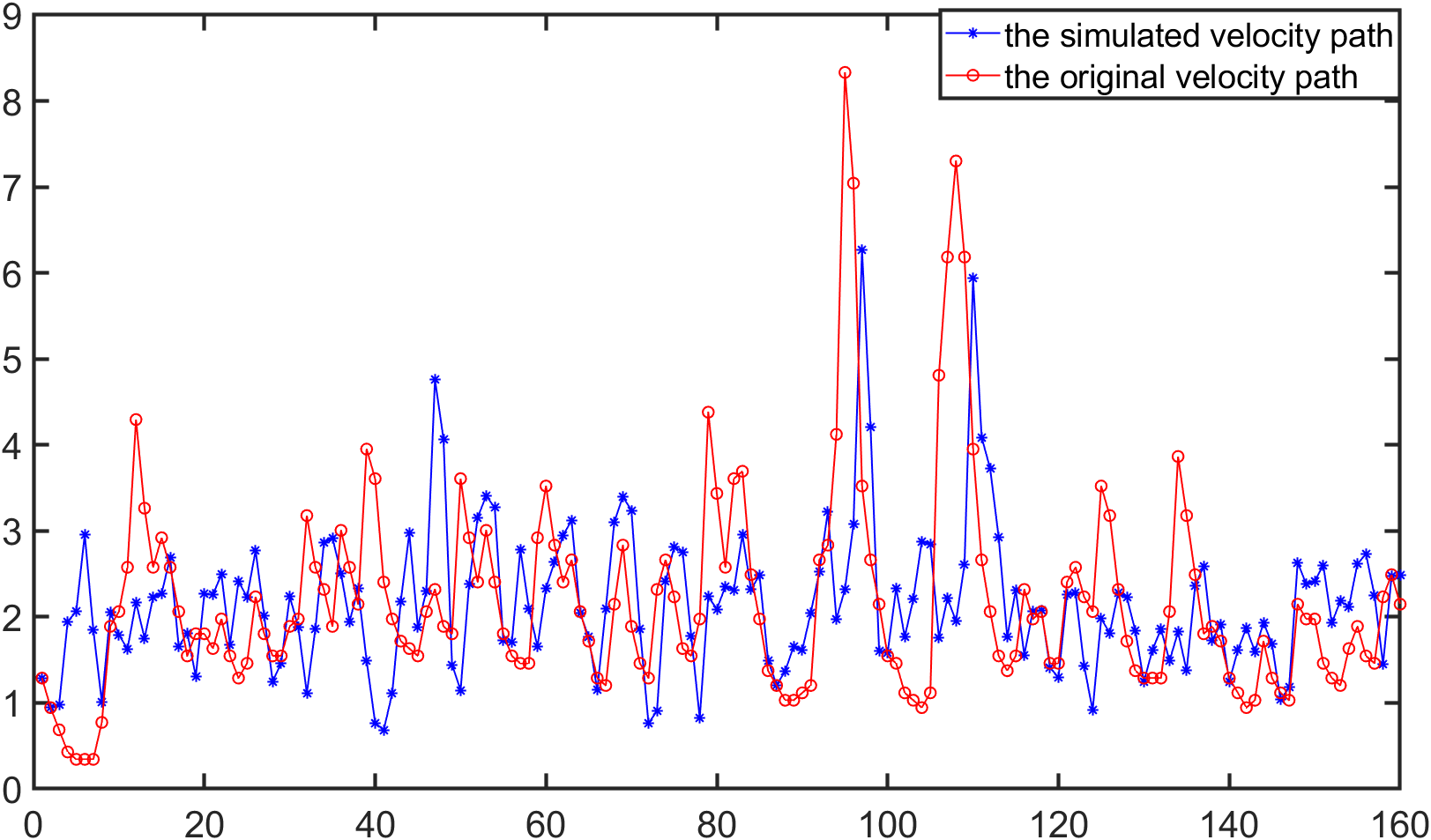}
}
\quad
\subfigure[]{
\includegraphics[width=5cm]{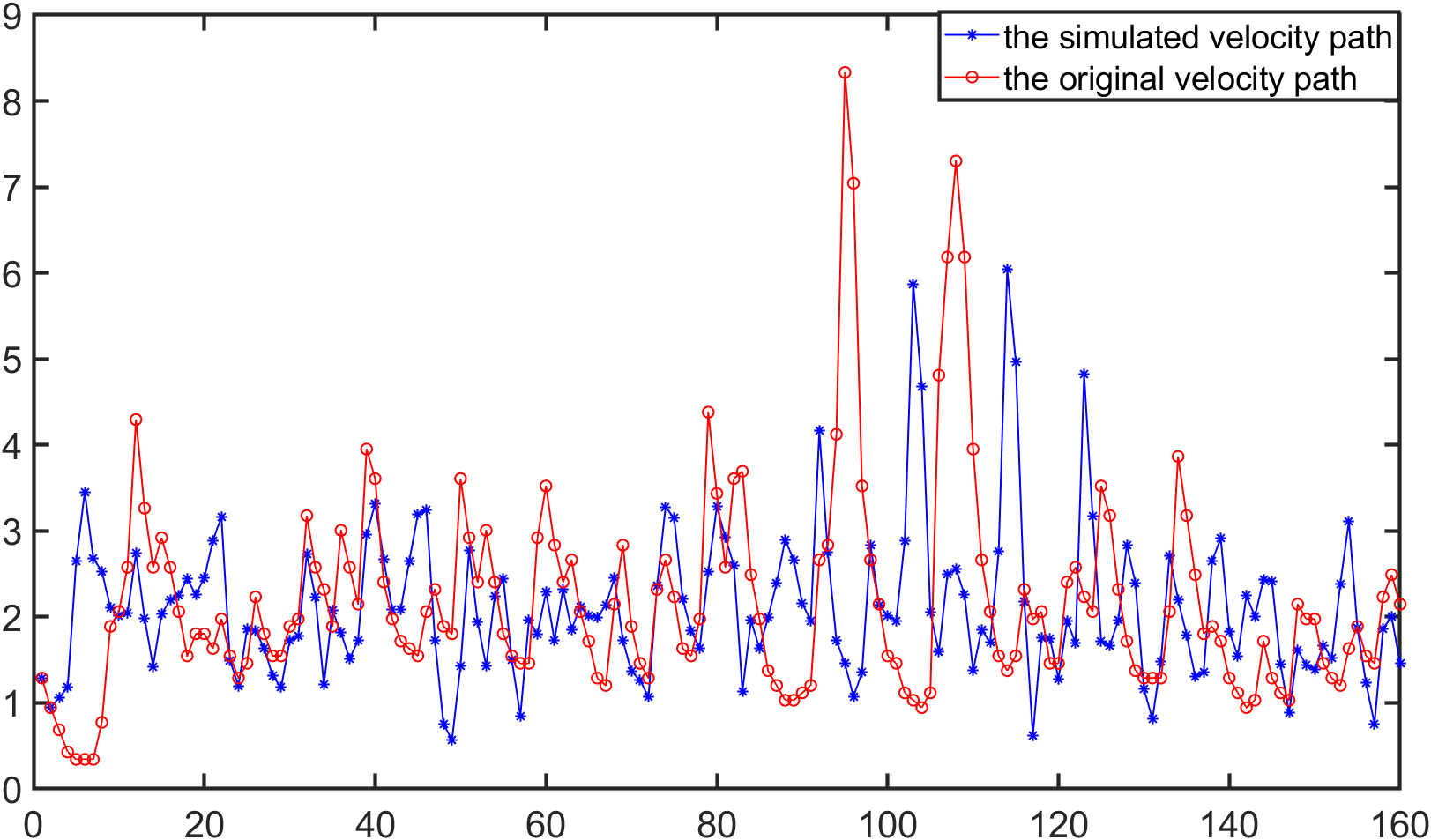}
}
\quad
\subfigure[]{
\includegraphics[width=5cm]{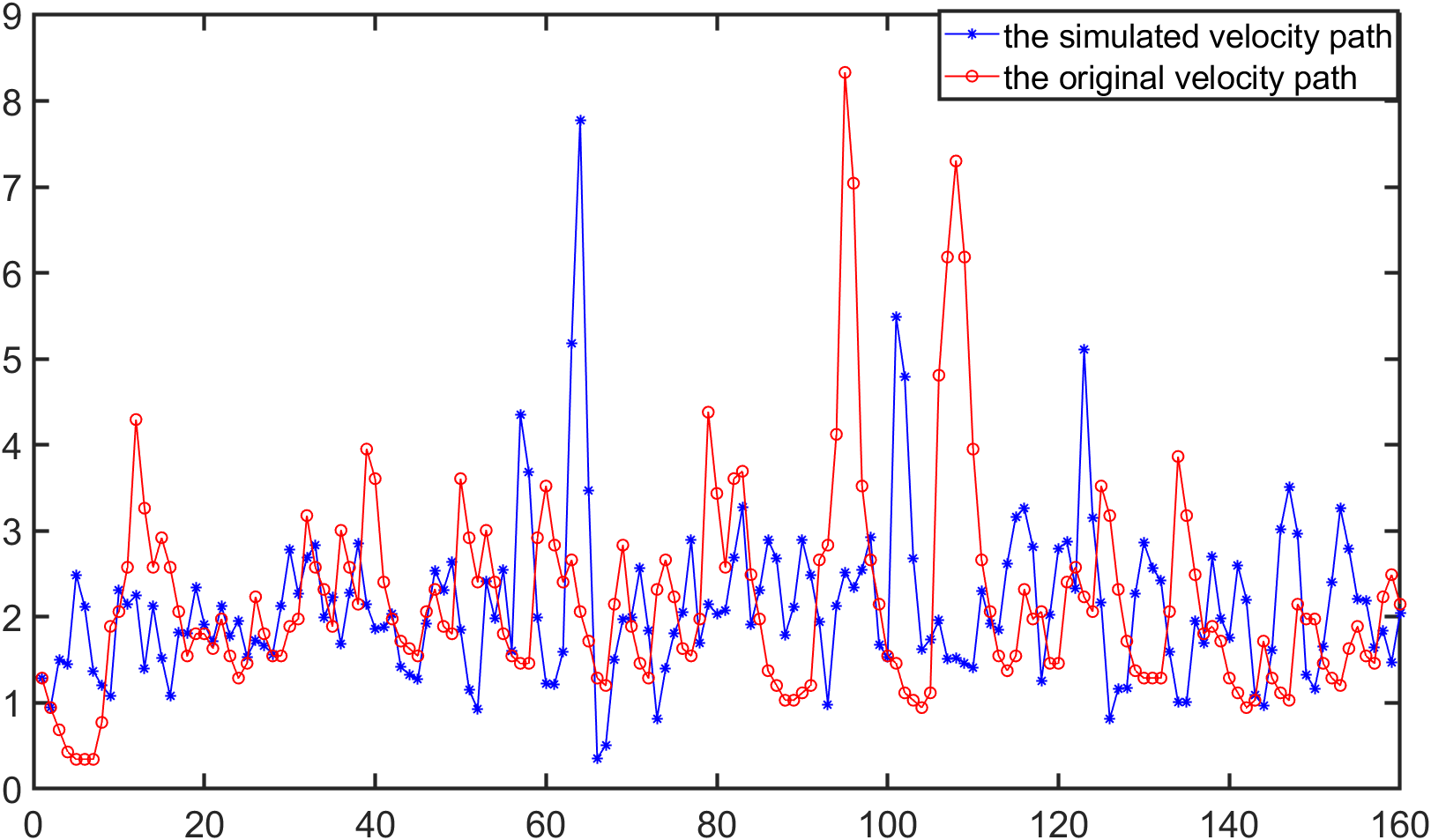}
}
\quad
\subfigure[]{
\includegraphics[width=5cm]{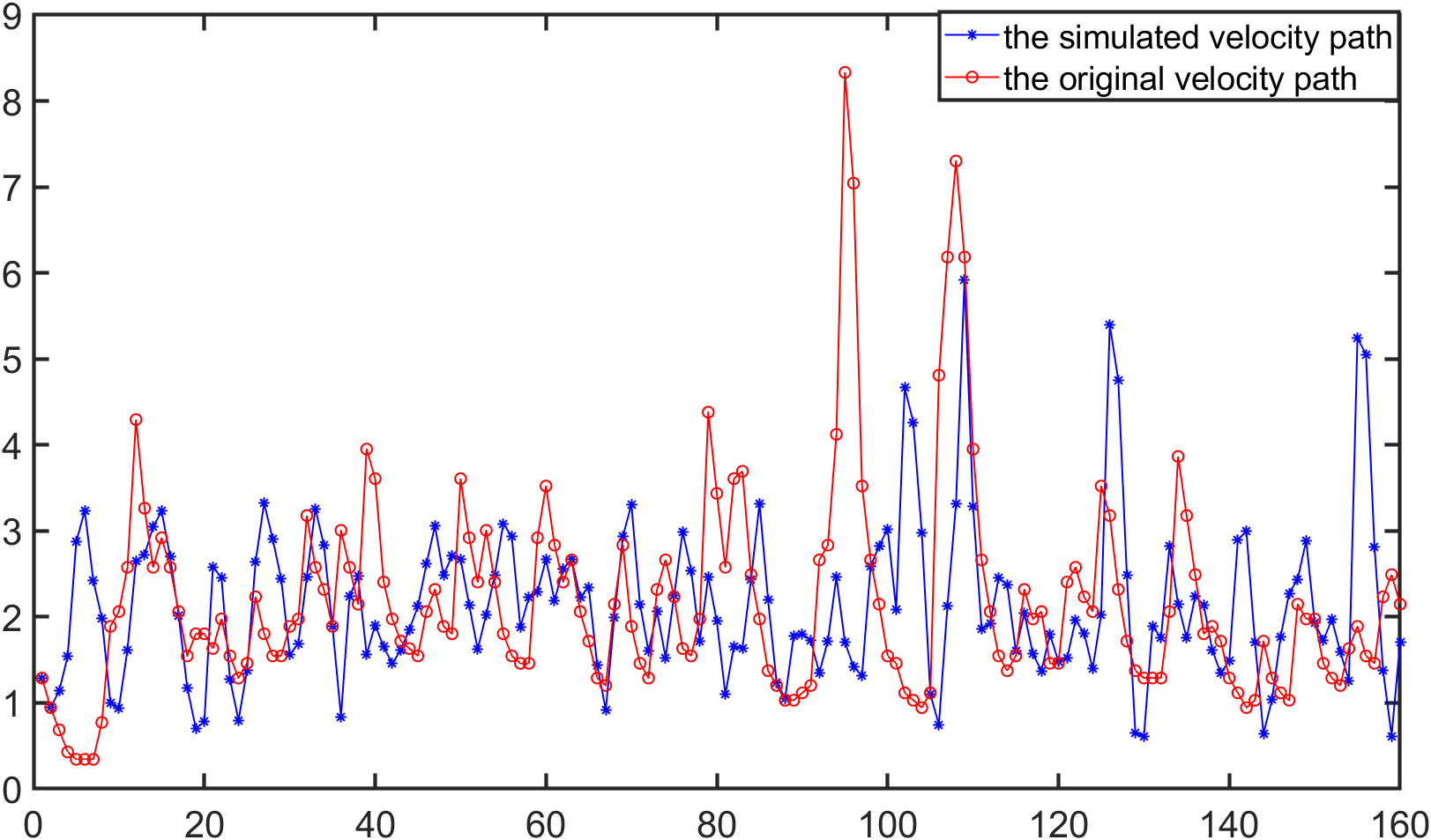}
}
\caption{The simulated velocity under the distribution $F_{u2}(v, 0.9683, (0.5470, 3.5392), (5.4610, 8.5577))$.}\label{twoun2}
\end{figure}

\begin{figure}[H]
\centering
\subfigure[]{
\includegraphics[width=5cm]{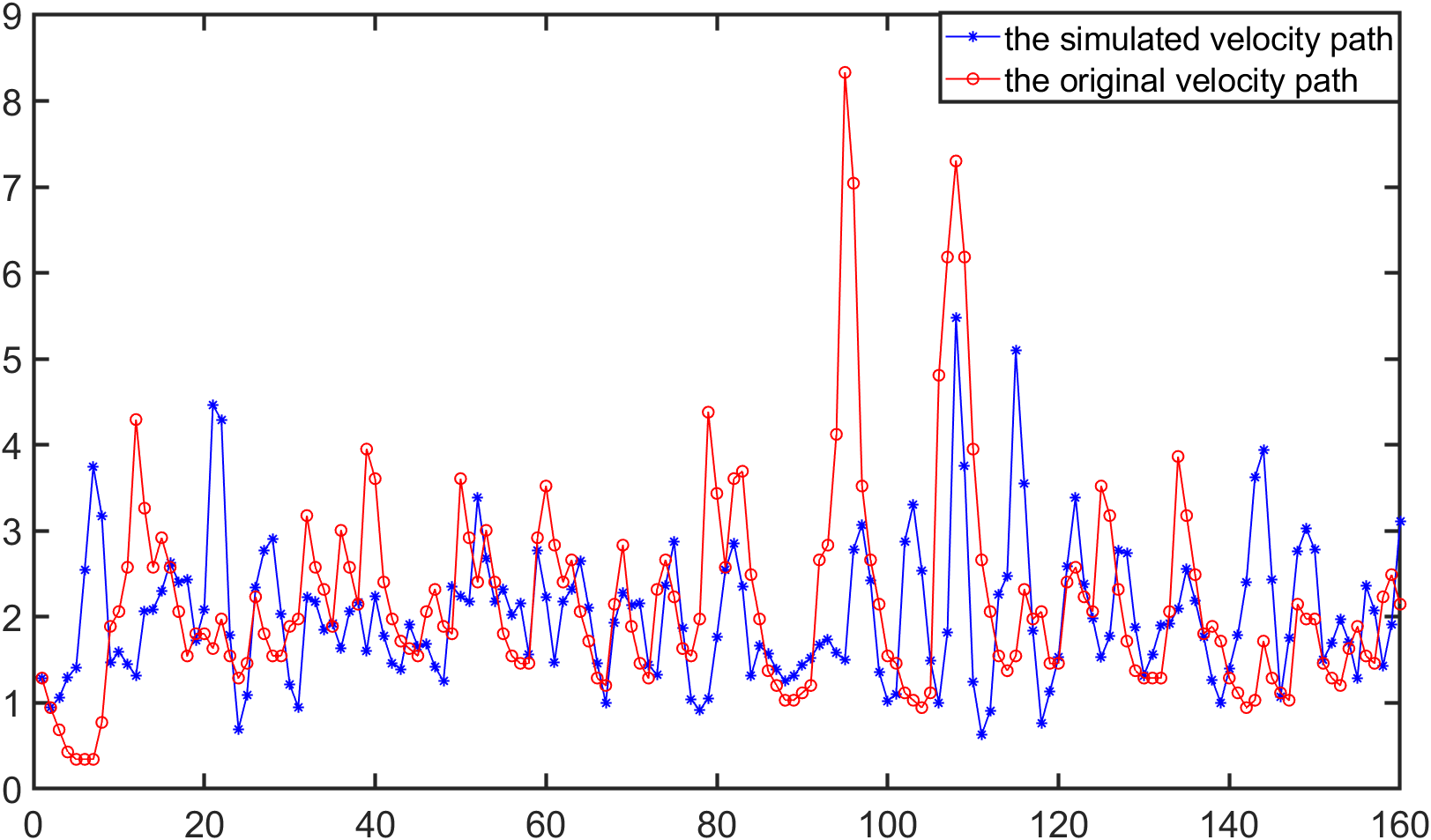}
}
\quad
\subfigure[]{
\includegraphics[width=5cm]{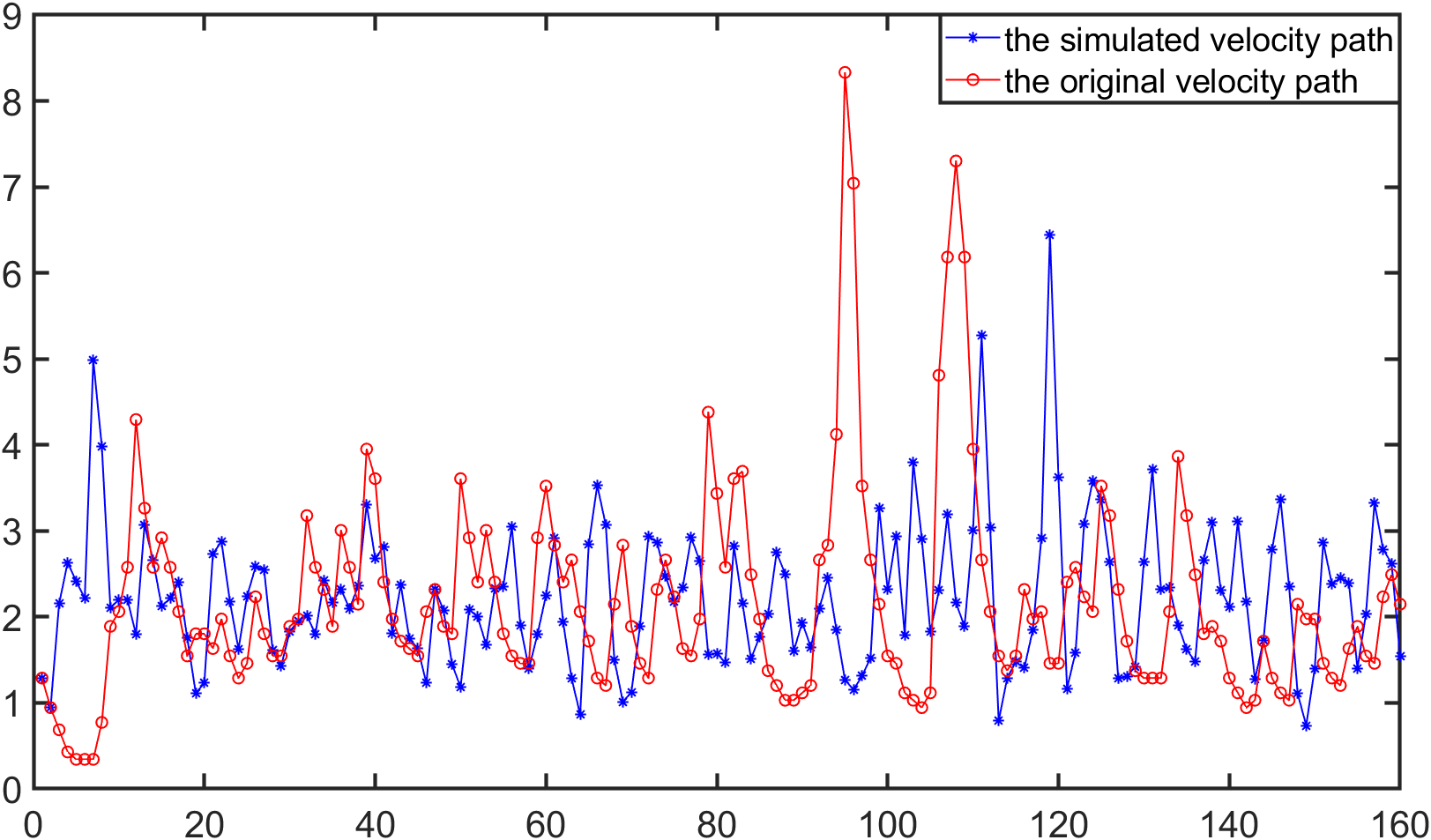}
}
\quad
\subfigure[]{
\includegraphics[width=5cm]{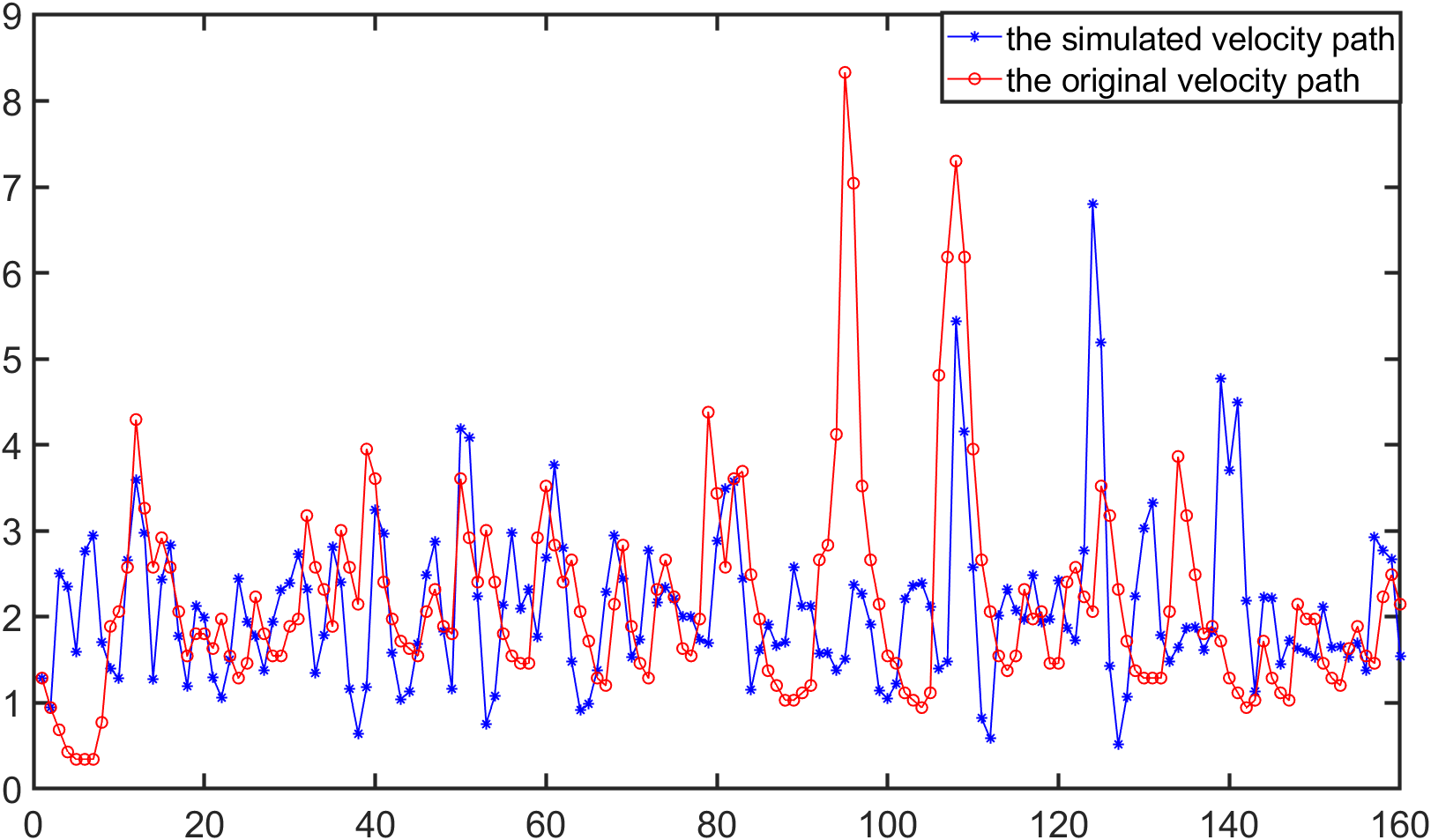}
}
\quad
\subfigure[]{
\includegraphics[width=5cm]{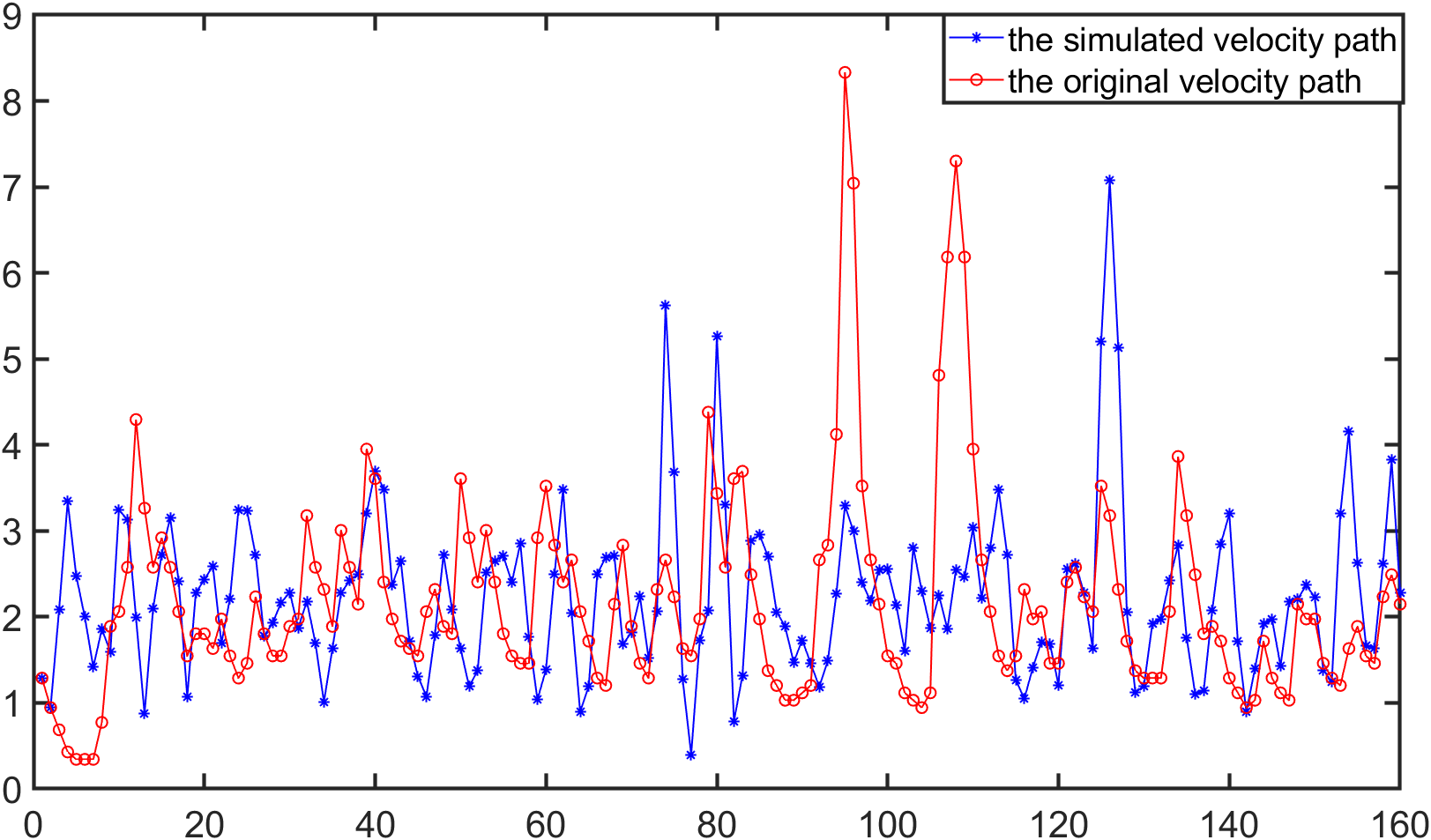}
}
\caption{The simulated velocity under $F_{t2}(v, 0.9683, (0.6740, 0.9752, 4.4800), (5.1118, 6.5112, 9.4053))$.}\label{twotr2}
\end{figure}

In this subsection, we propose more precise distributions of velocity $v$ to preserve the expectations and variances of the velocity sampling in two parts, which are $(m_1, \sigma_1^2)=(2.0431, 0.7461)$ and $(m_2, \sigma_2^2)=(7.0094, 0.7991)$ respectively. This means that we find the parameters of the superposition of two uniform, two Wigner semicircle or two triangular distributions ($F_{u2}, F_{W2}, F_{t2}$), thus reproducing $(m_1, \sigma_1^2)$ and $(m_2, \sigma_2^2)$.
%that is, the expectations $m_i$ and variances $\sigma_i^2$ of the distributions preserve the sampling expectation and variance, respectively.
Then  the behavior of a falling ball is simulated by the algorithm (\ref{newmain}) and the results are shown in Figures \ref{twosc2}, \ref{twoun2} and \ref{twotr2}.

To get the parameters $(a_i, b_i, c_i)$ of the triangular distribution keeping the expectation and variance of the samples, we especially add the condition of keeping the geometrical center of velocity.
\par
These more precise distributions of velocity $v$ here reproducing the expectation and variance.
The results of numerical examples have showed that the algorithm (\ref{newmain}) combining the Newton-Stokes algorithm and the soft Lasso's algorithm can satisfactorily represent the pattern of a ball falling in a non-Newtonian fluid.
The trajectory of the motion of the falling ball is random and  not experimentally reproducible.

\subsection{Pattern preserving}
\begin{figure}[H]
\centering
\subfigure[]{
\includegraphics[width=3.5cm]{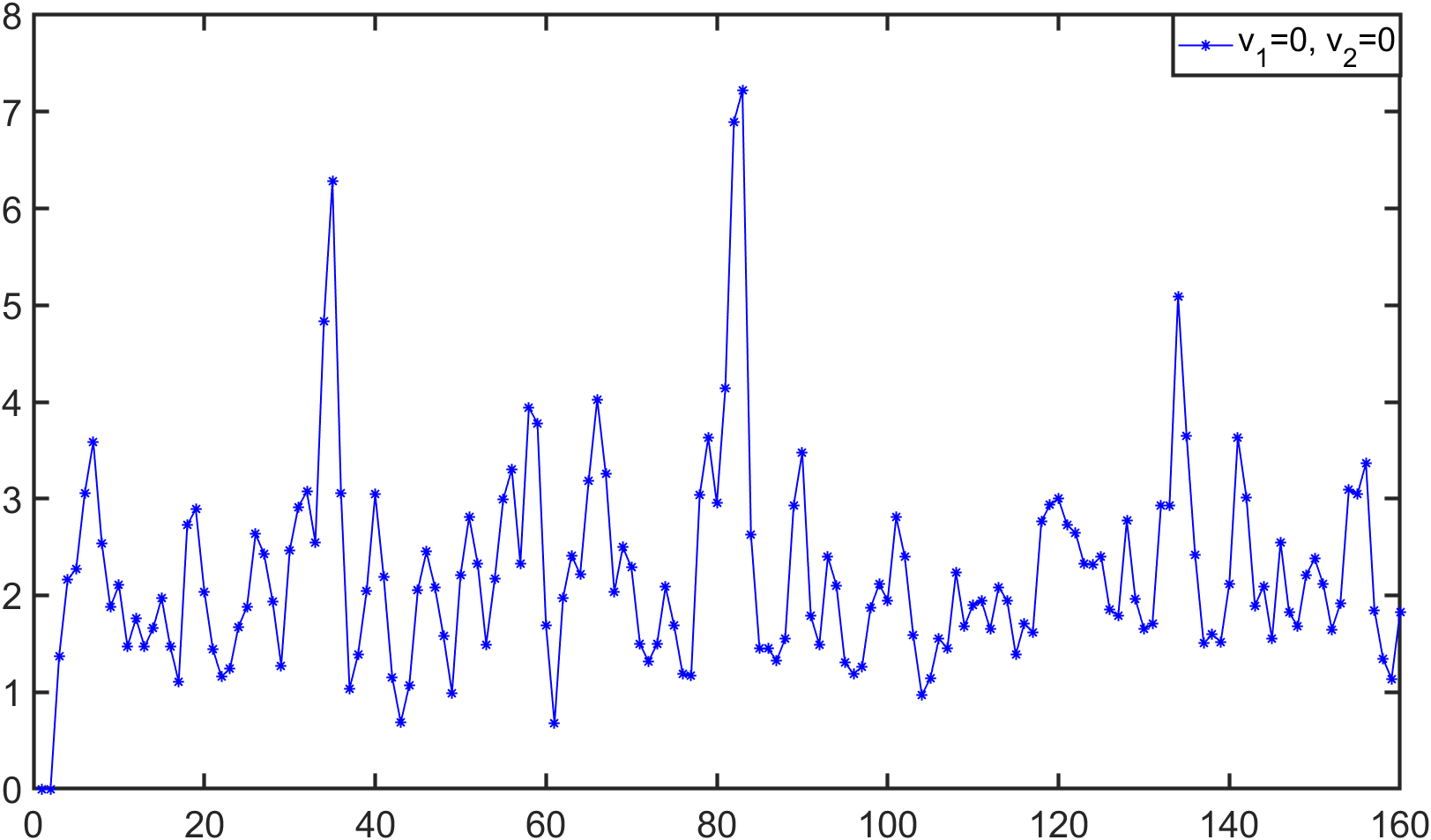}
}
\quad
\subfigure[]{
\includegraphics[width=3.5cm]{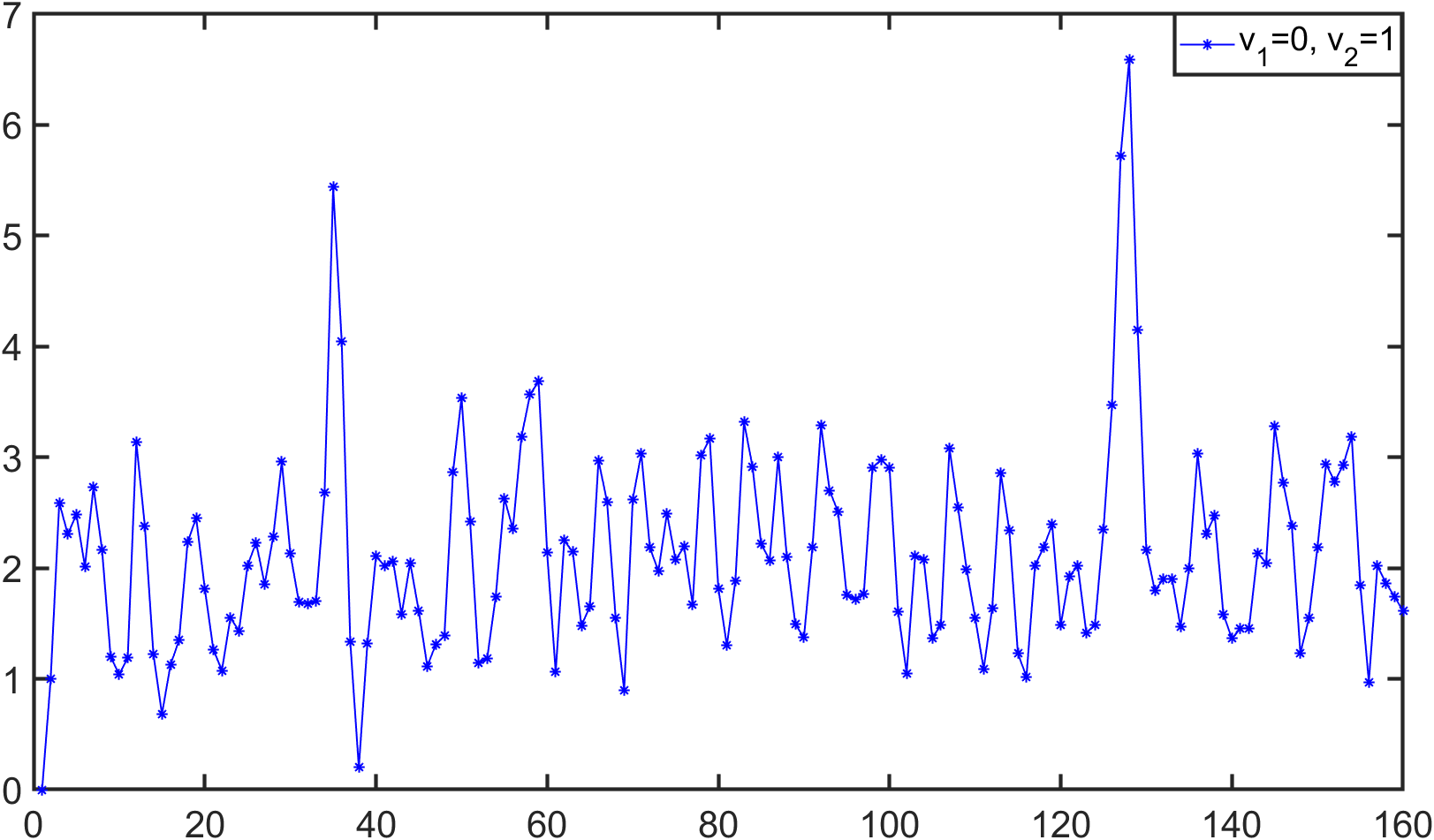}
}
\quad
\subfigure[]{
\includegraphics[width=3.5cm]{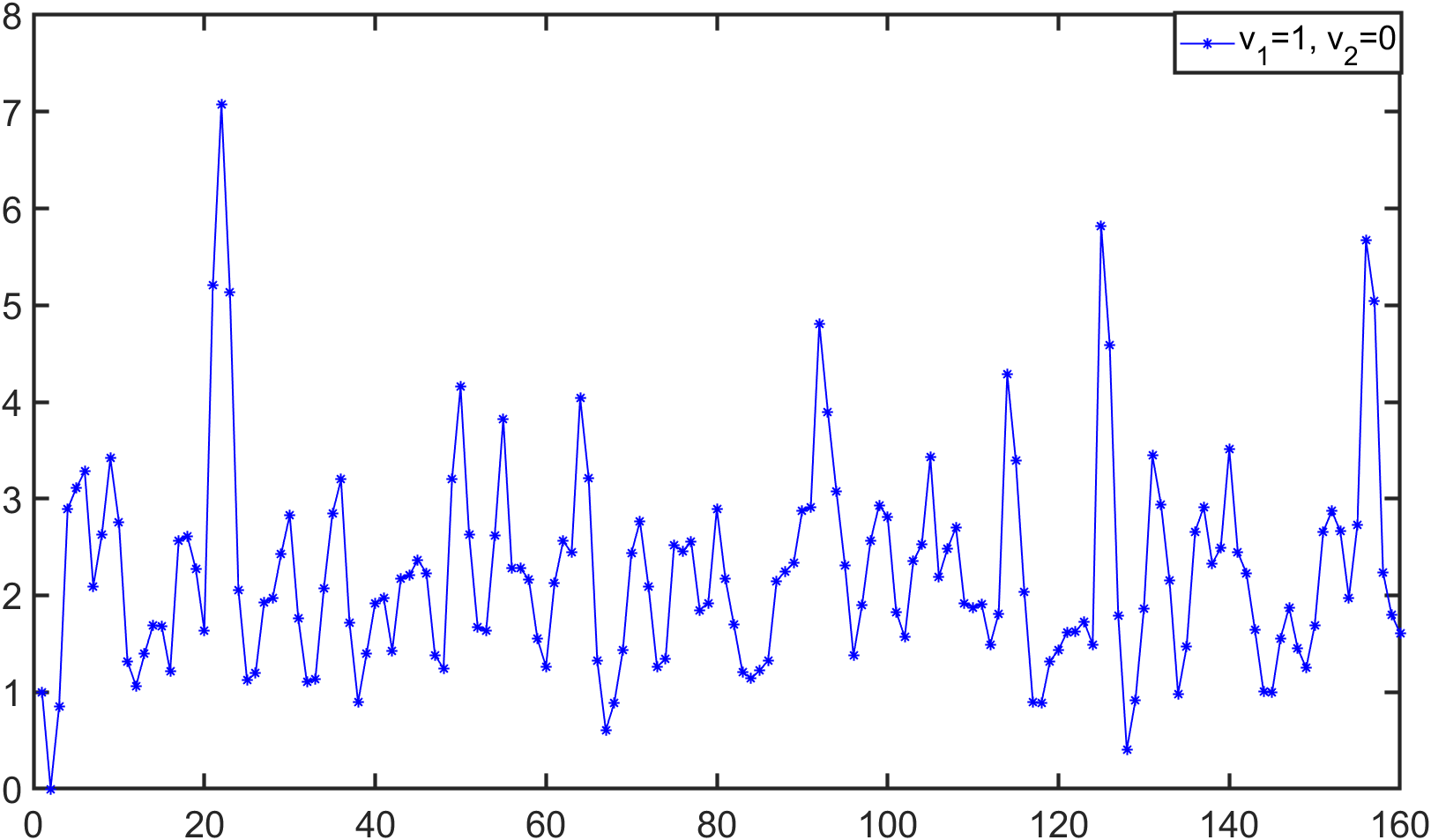}
}
\quad
\subfigure[]{
\includegraphics[width=3.5cm]{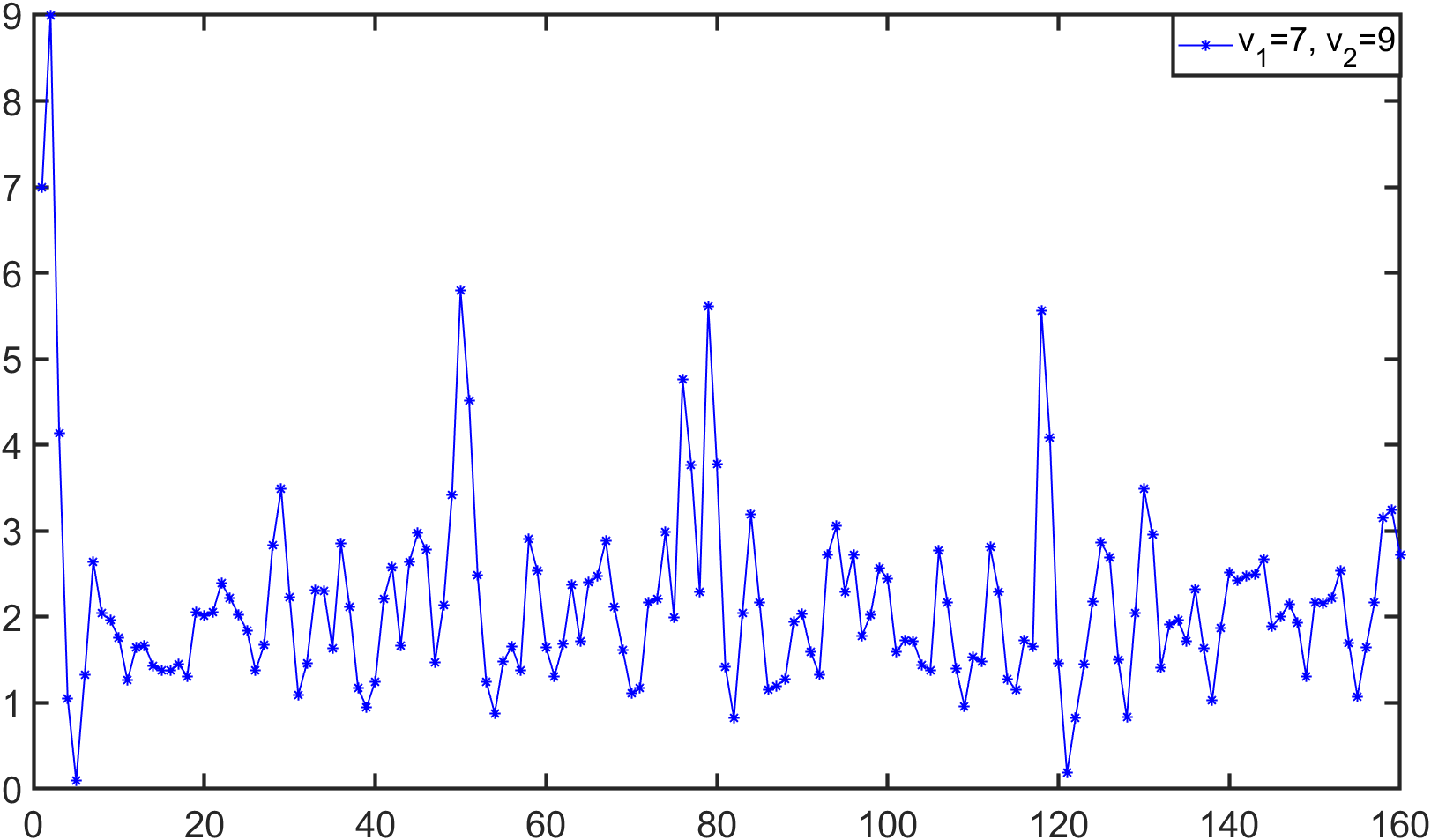}
}
\quad
\subfigure[]{
\includegraphics[width=3.5cm]{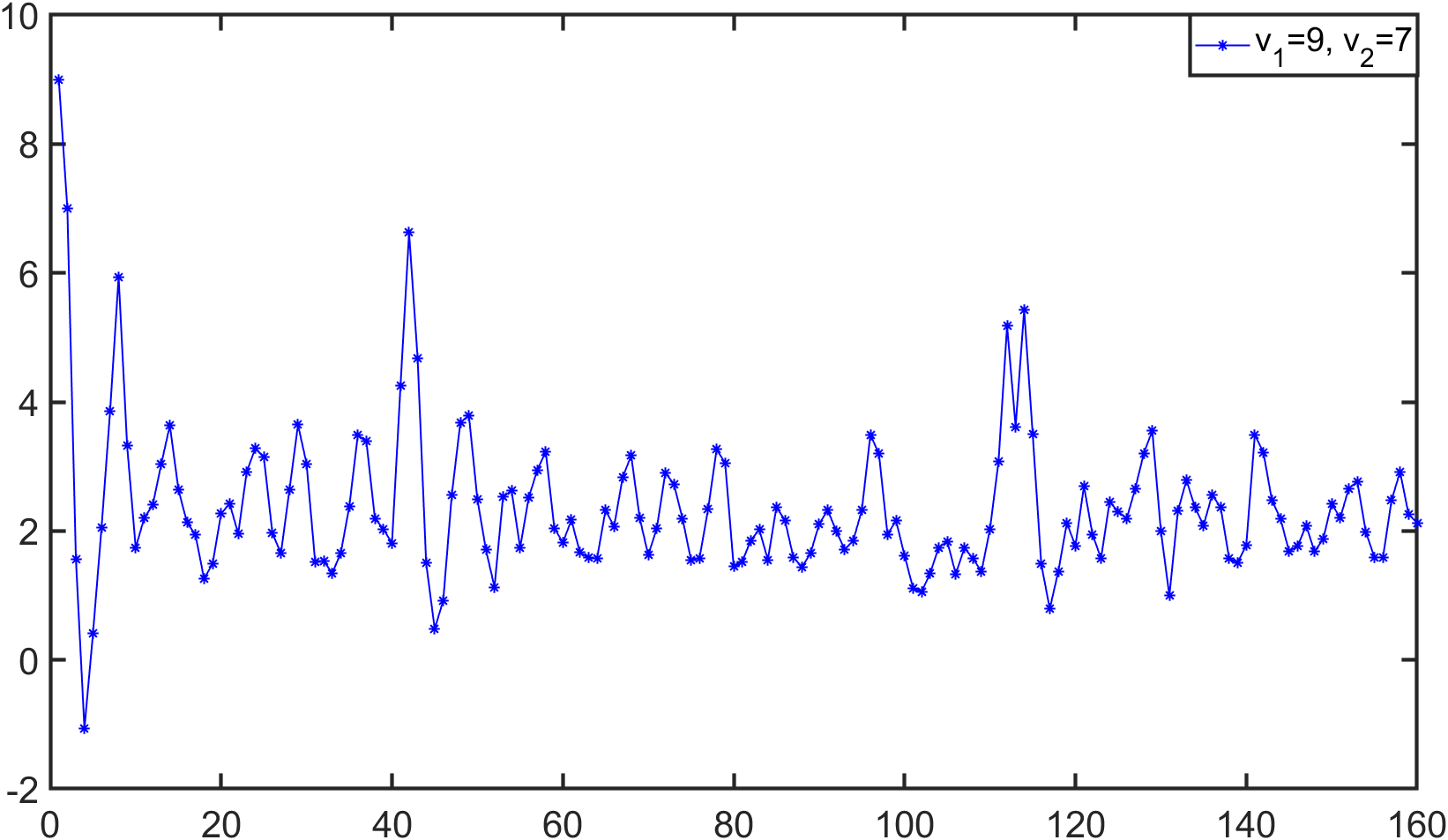}
}
\quad
\subfigure[]{
\includegraphics[width=3.5cm]{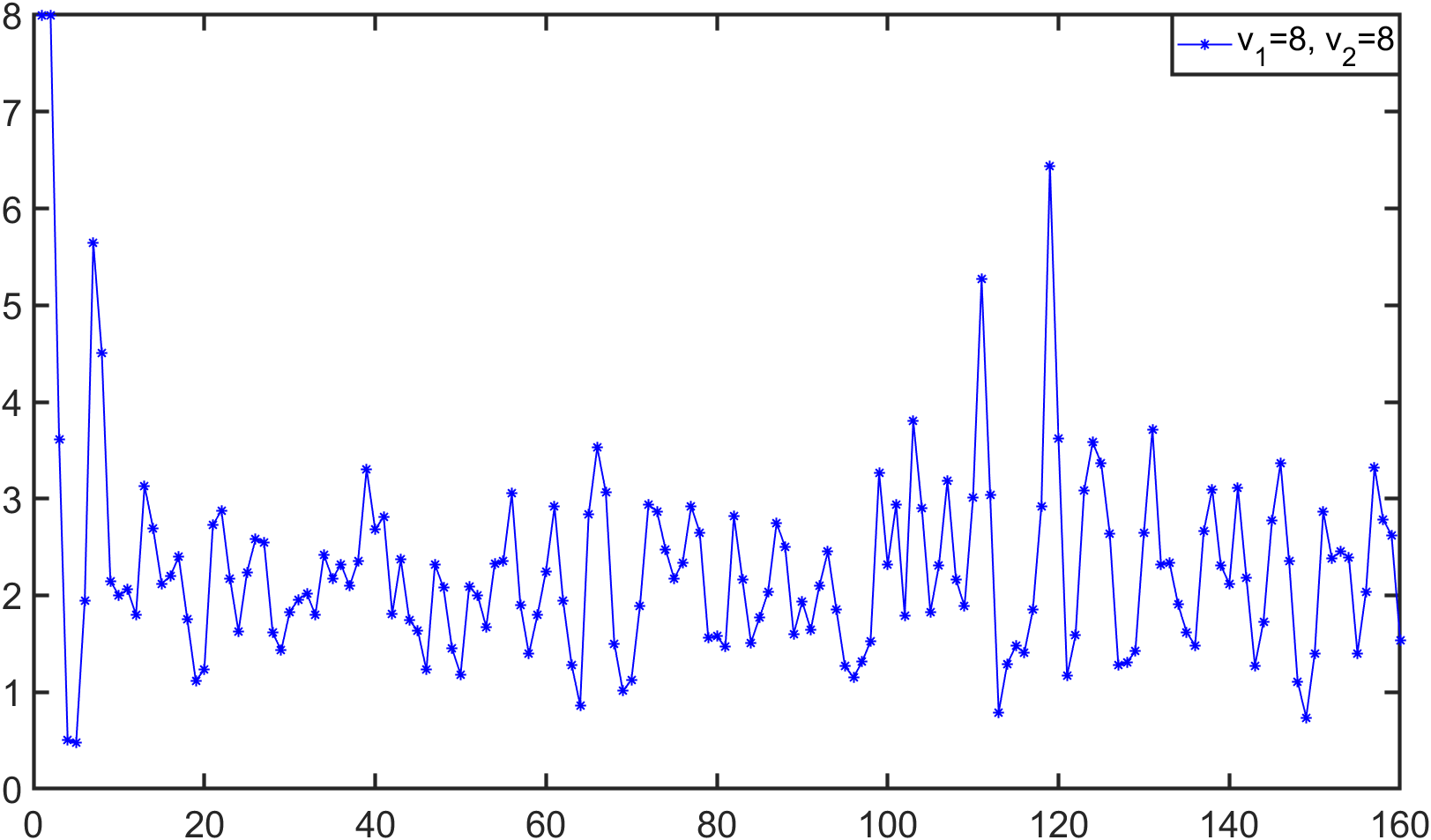}
}
\quad
\subfigure[]{
\includegraphics[width=3.5cm]{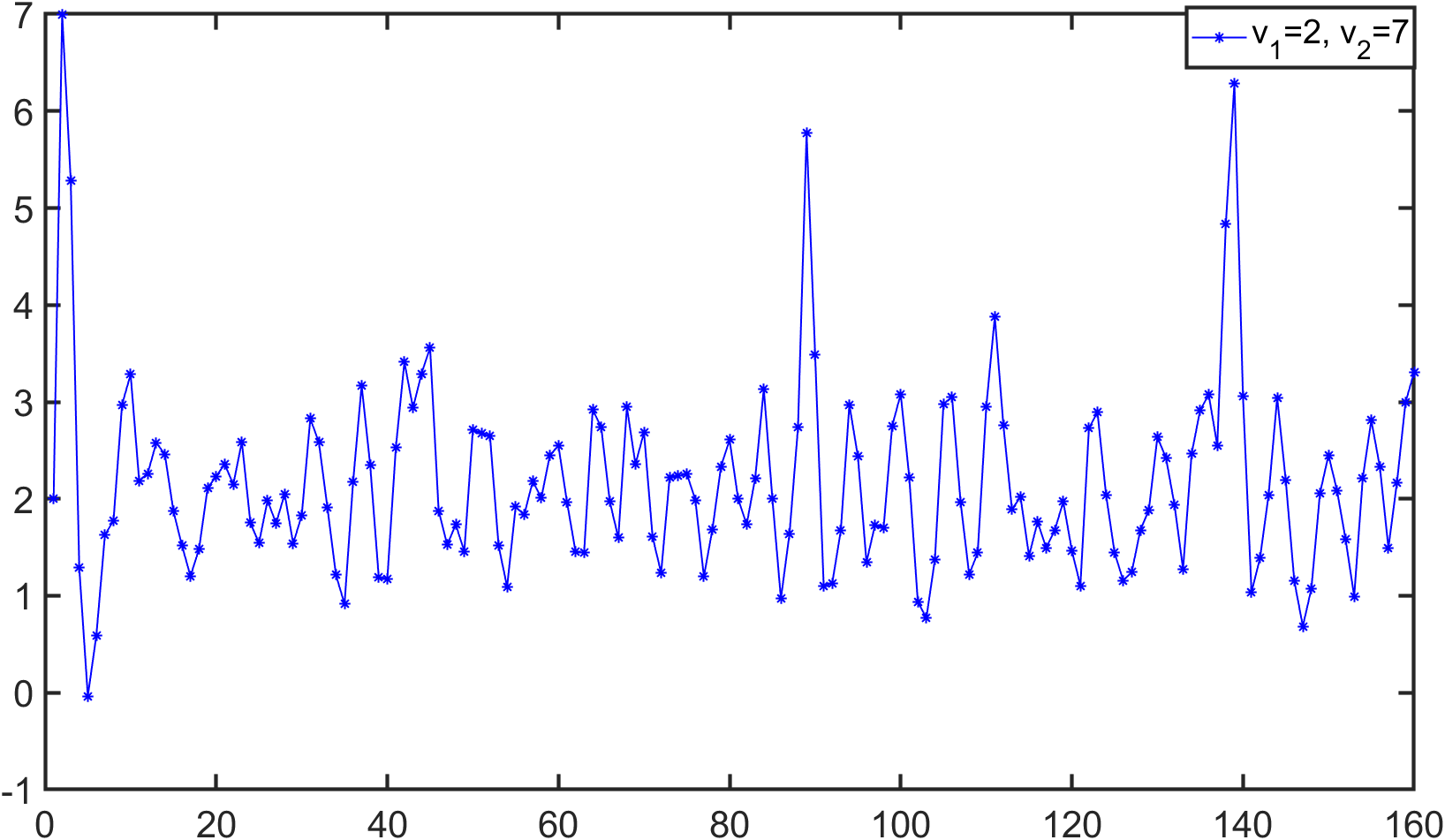}
}
\quad
\subfigure[]{
\includegraphics[width=3.5cm]{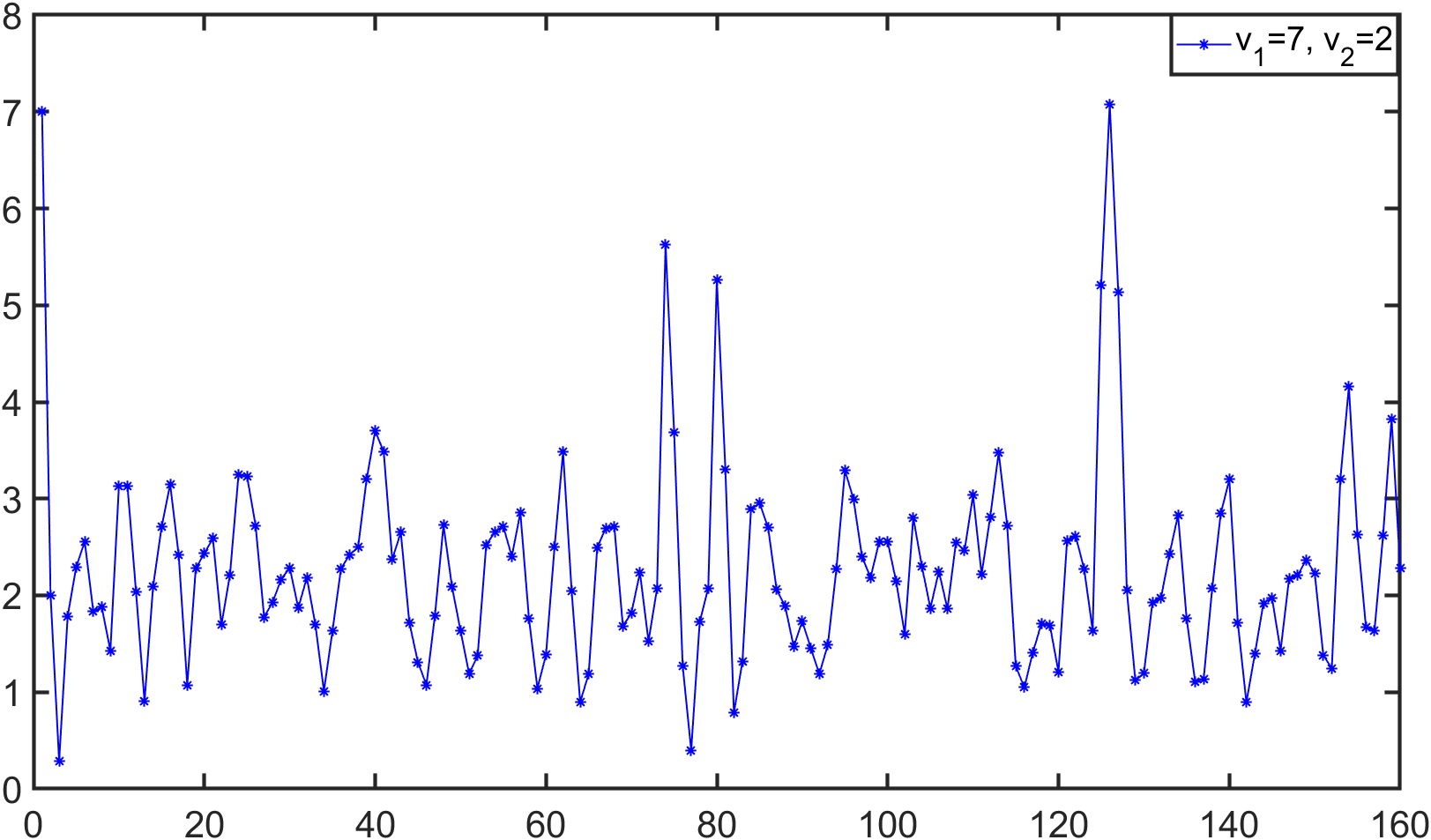}
}
\quad
\subfigure[]{
\includegraphics[width=3.5cm]{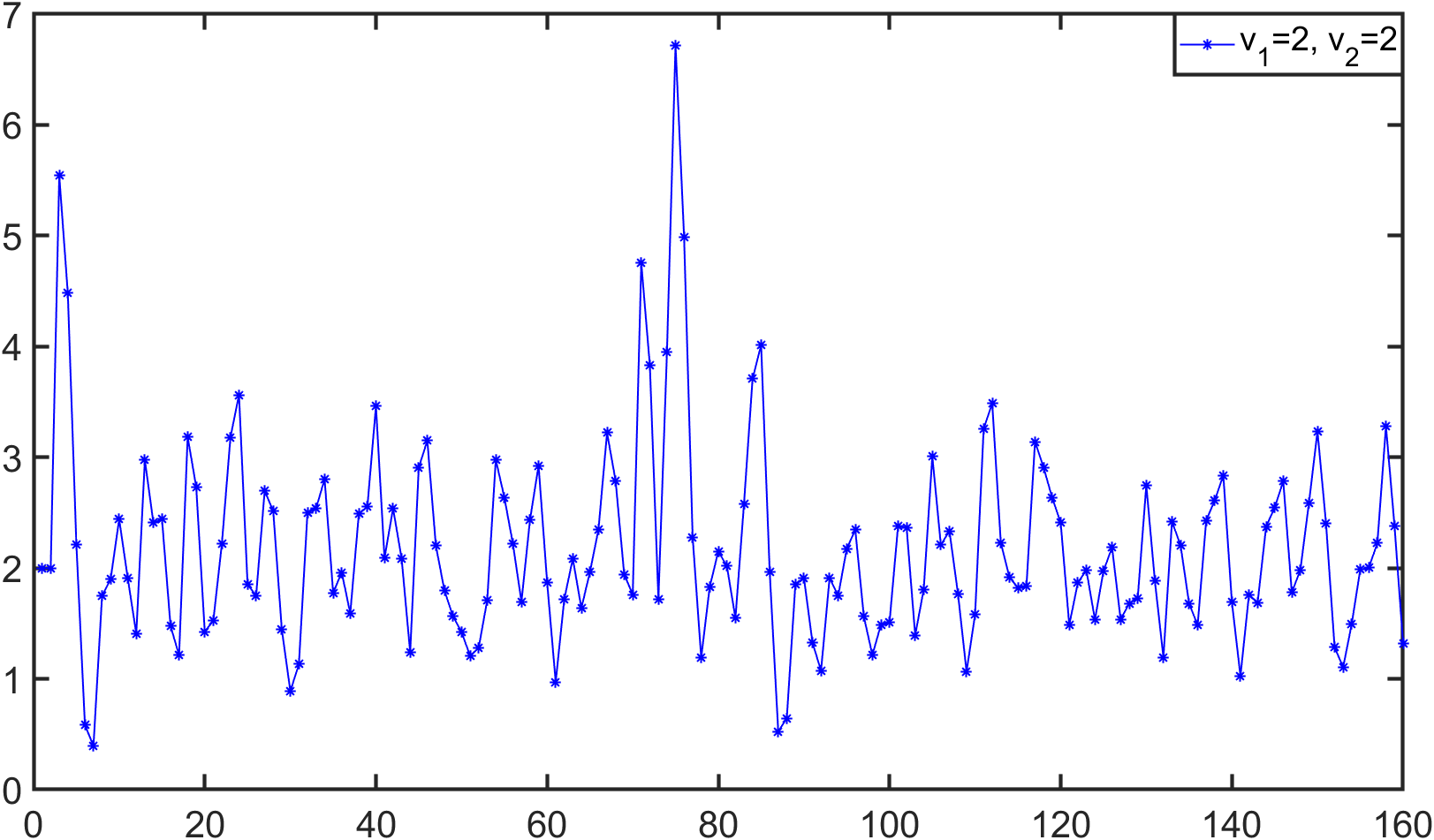}
}
\caption{Simulated trajectory of velocity by using different initial values.}\label{initialtwotr2}
\end{figure}

\begin{figure}[H]
\centering
\subfigure[]{
\includegraphics[width=3.5cm]{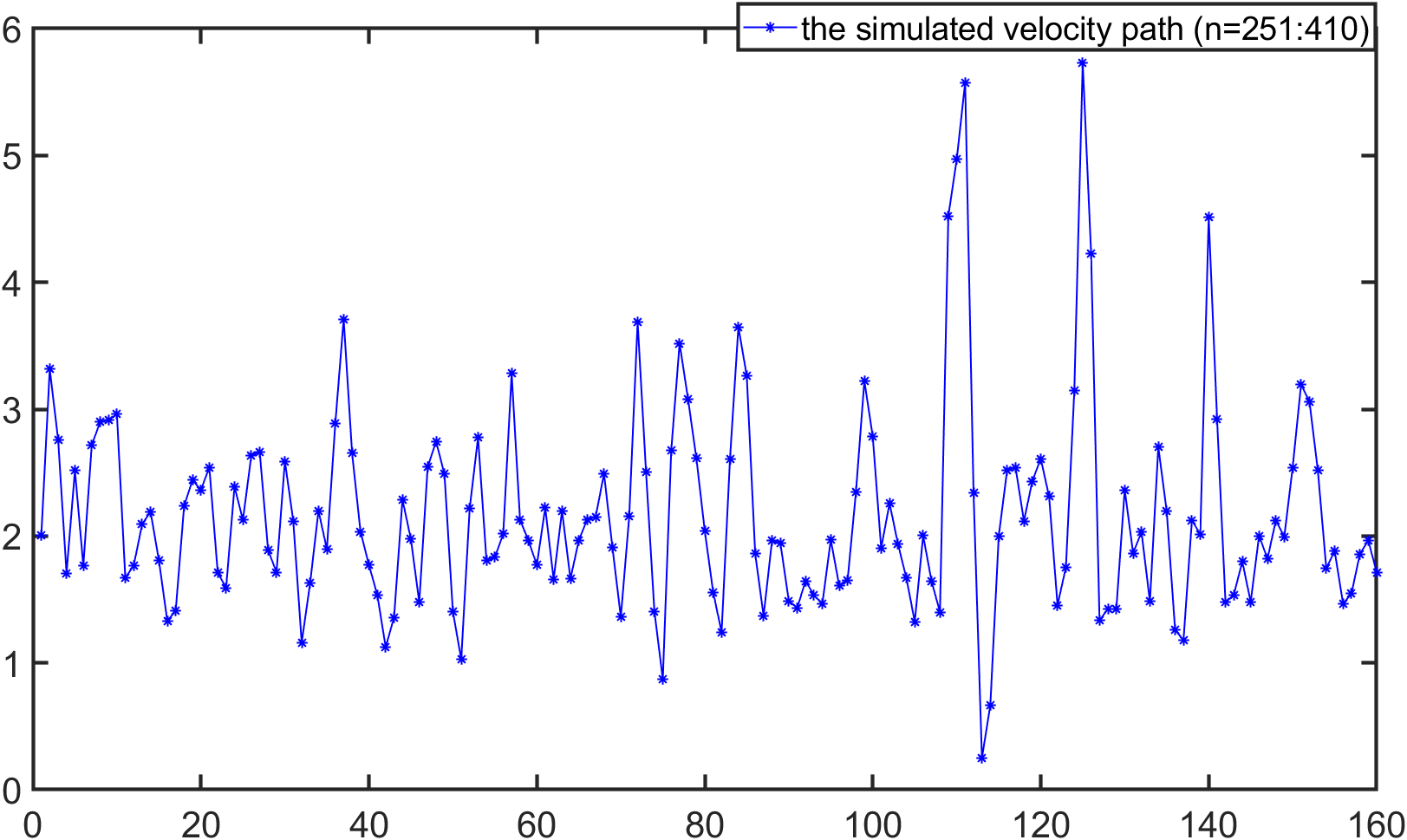}
}
\quad
\subfigure[]{
\includegraphics[width=3.5cm]{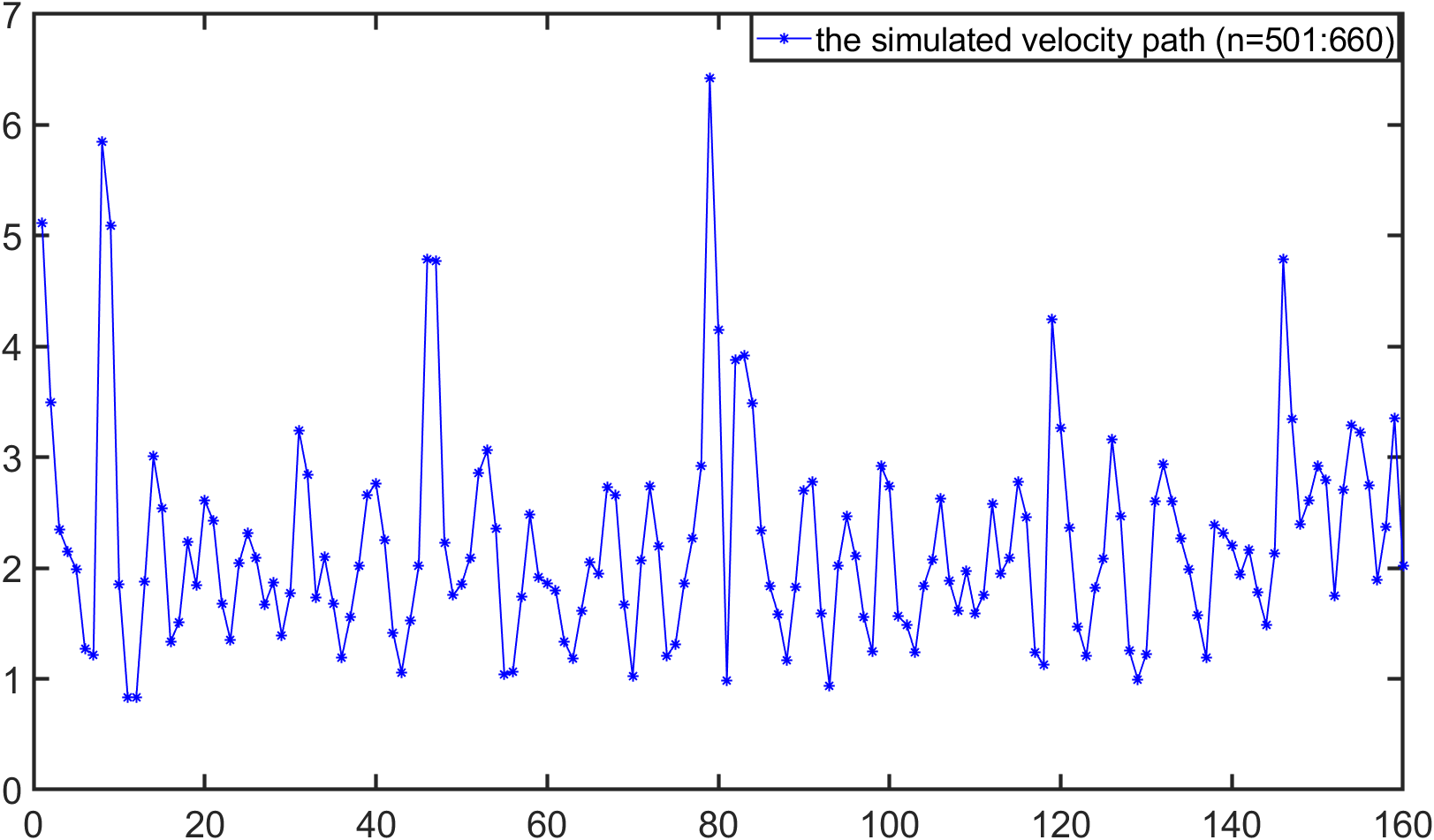}
}
\quad
\subfigure[]{
\includegraphics[width=3.5cm]{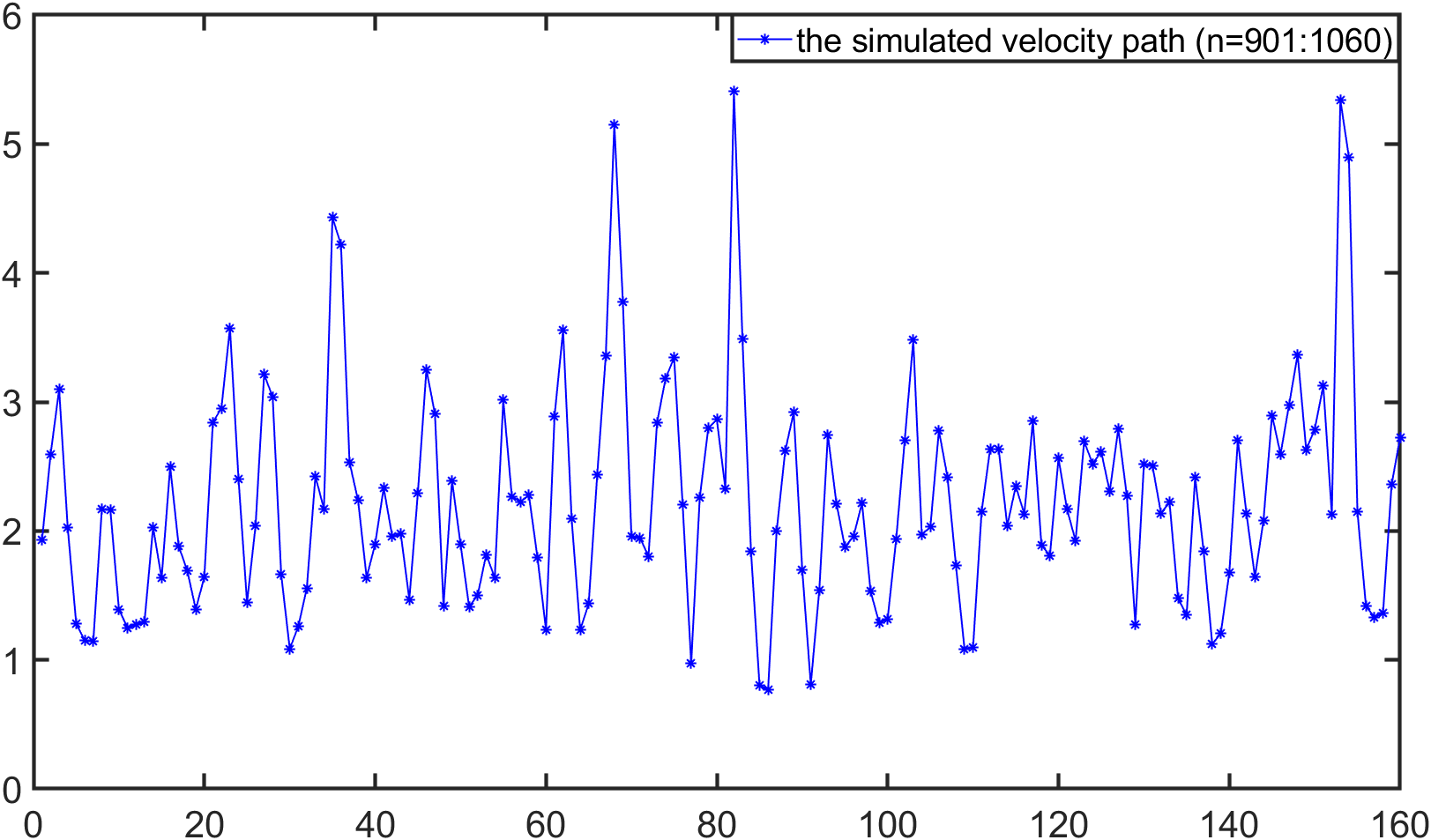}
}
\quad
\subfigure[]{
\includegraphics[width=3.5cm]{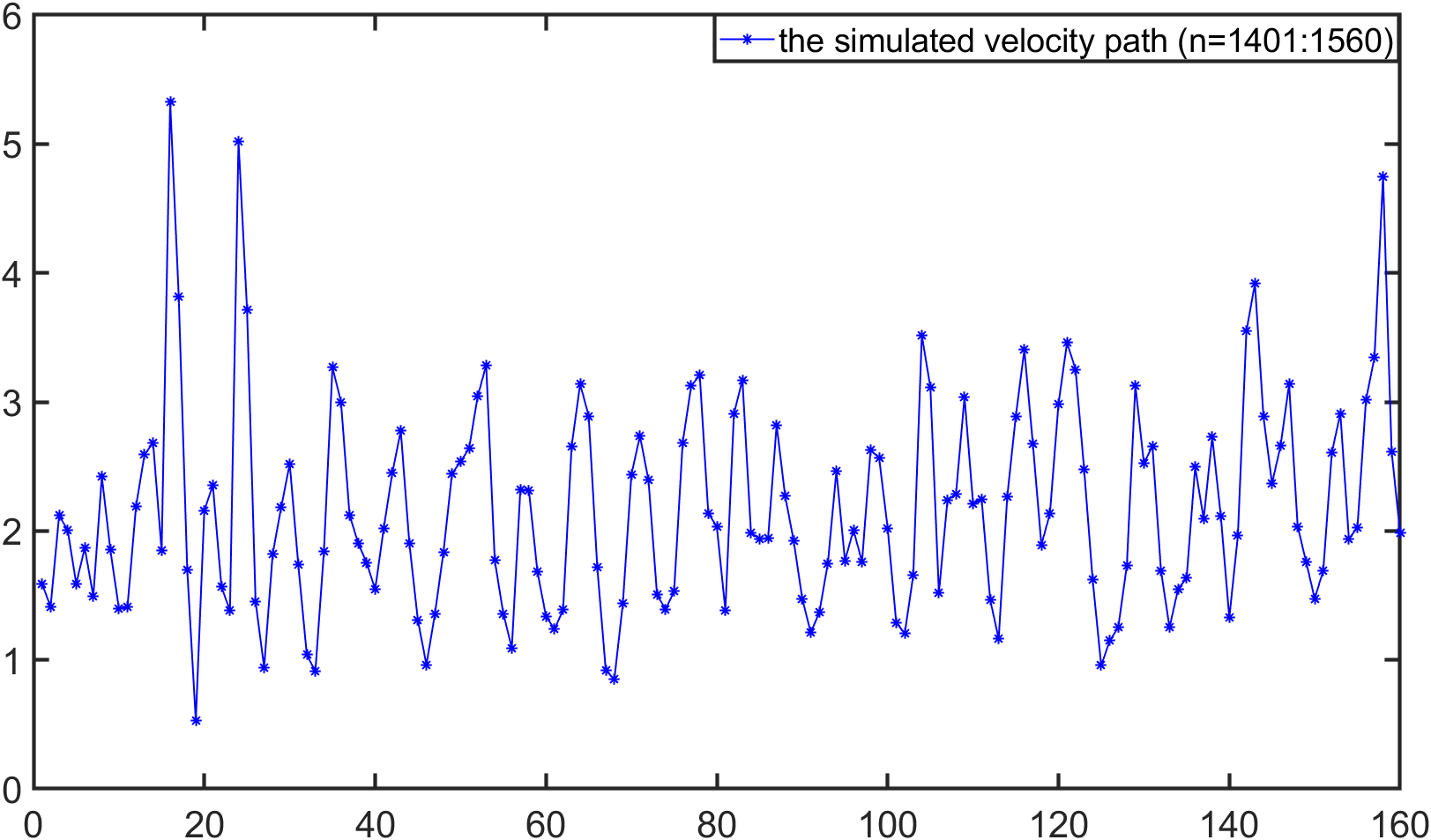}
}
\quad
\subfigure[]{
\includegraphics[width=3.5cm]{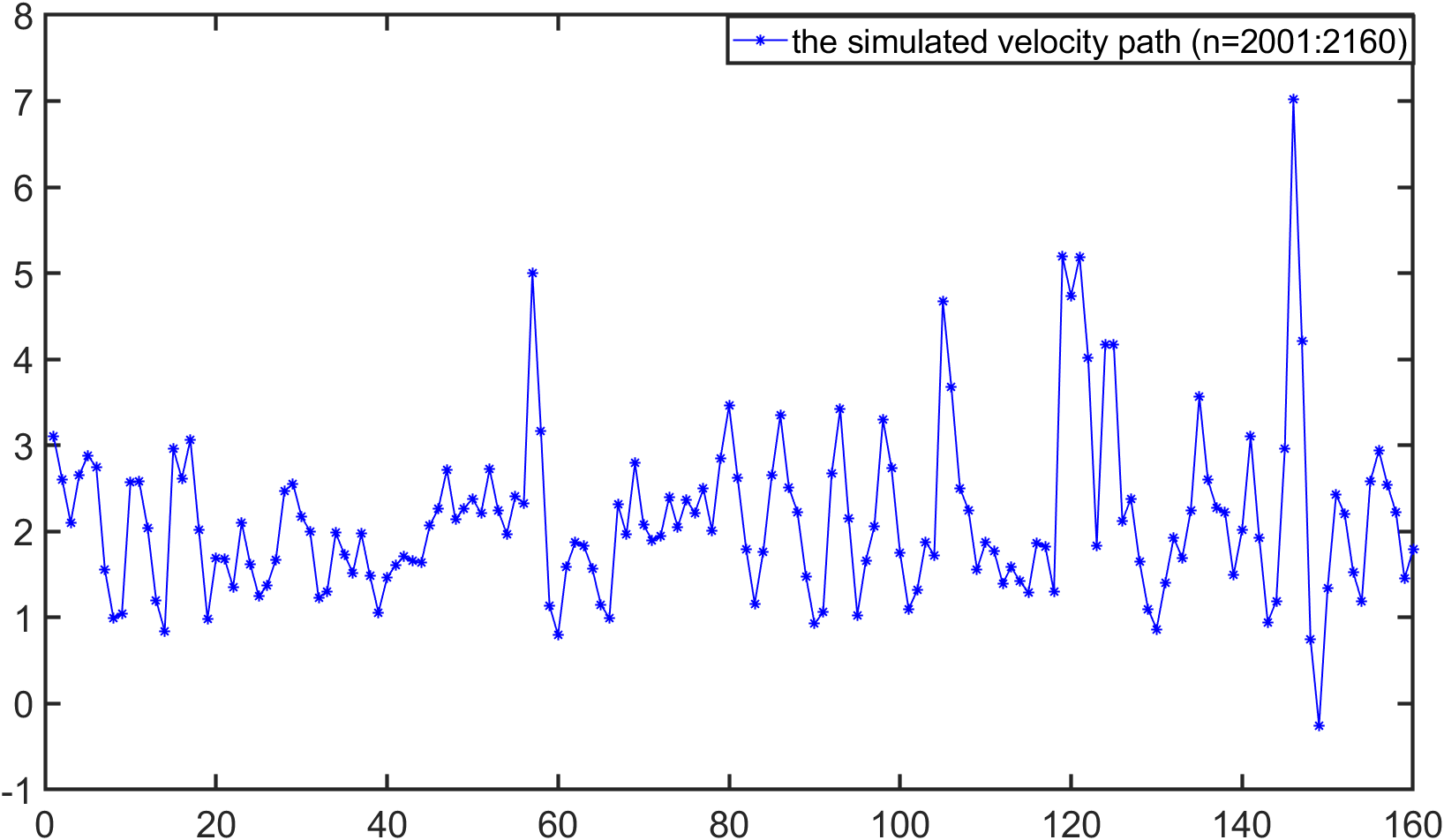}
}
\quad
\subfigure[]{
\includegraphics[width=3.5cm]{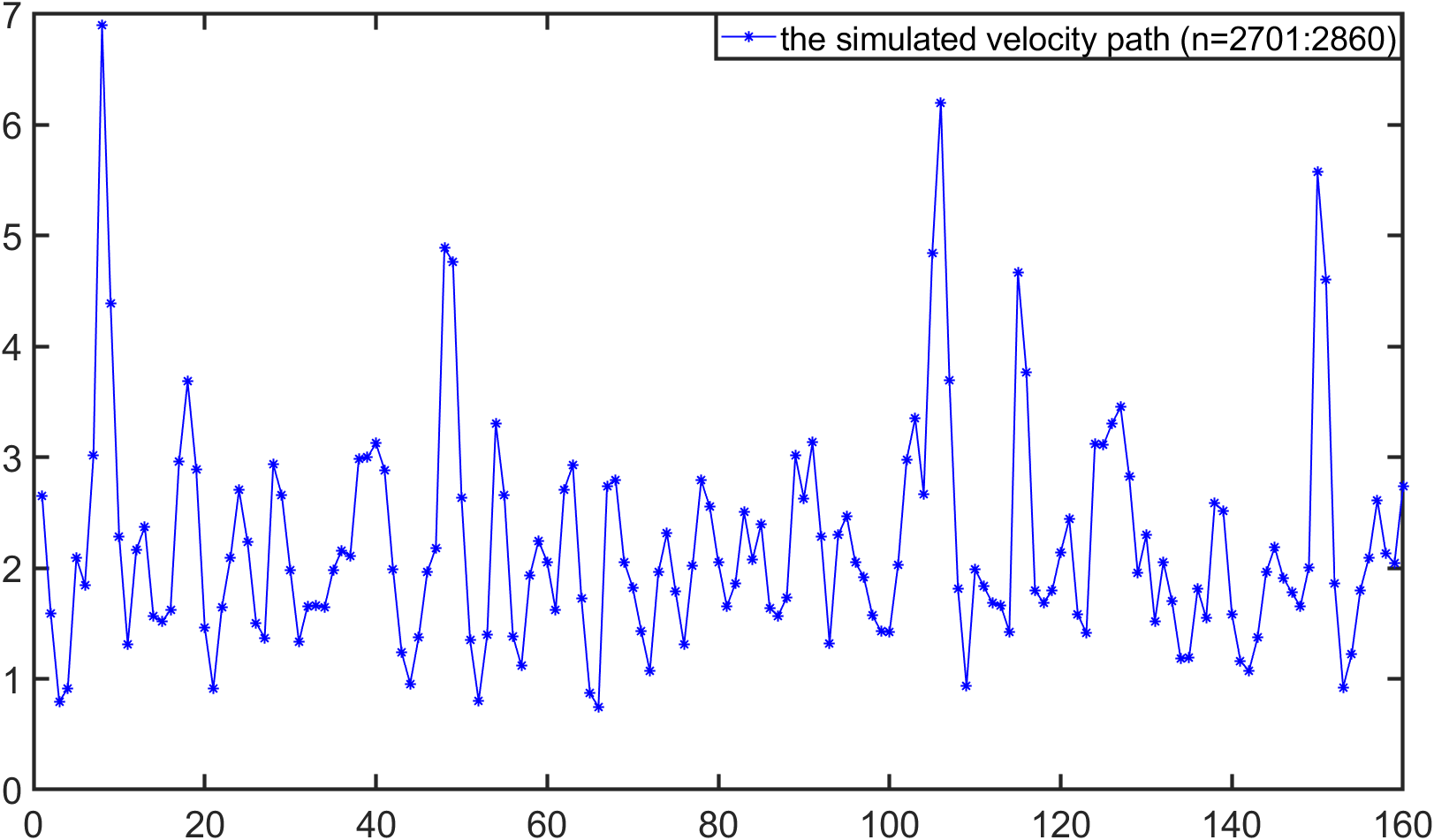}
}
\quad
\subfigure[]{
\includegraphics[width=3.5cm]{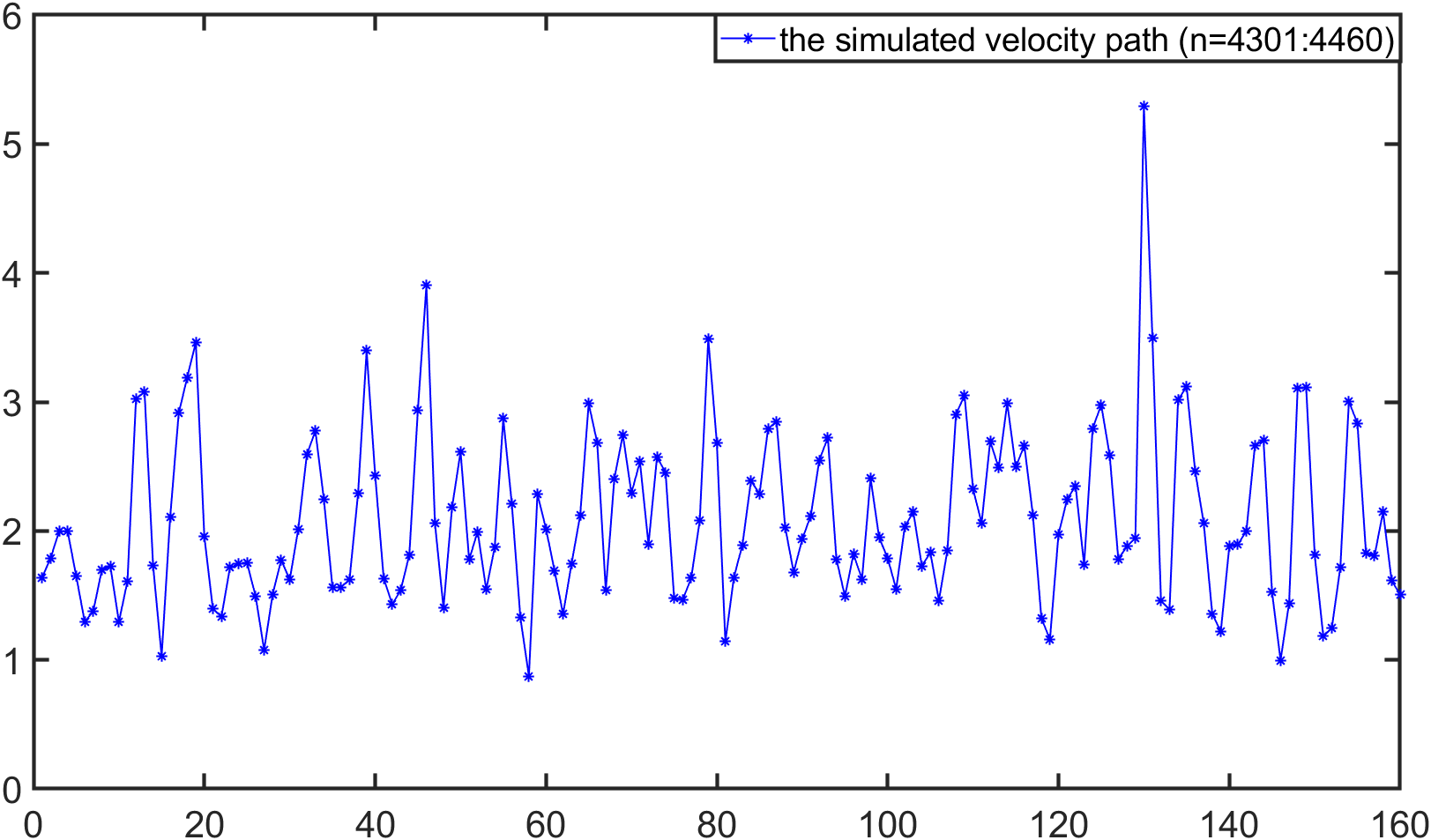}
}
\quad
\subfigure[]{
\includegraphics[width=3.5cm]{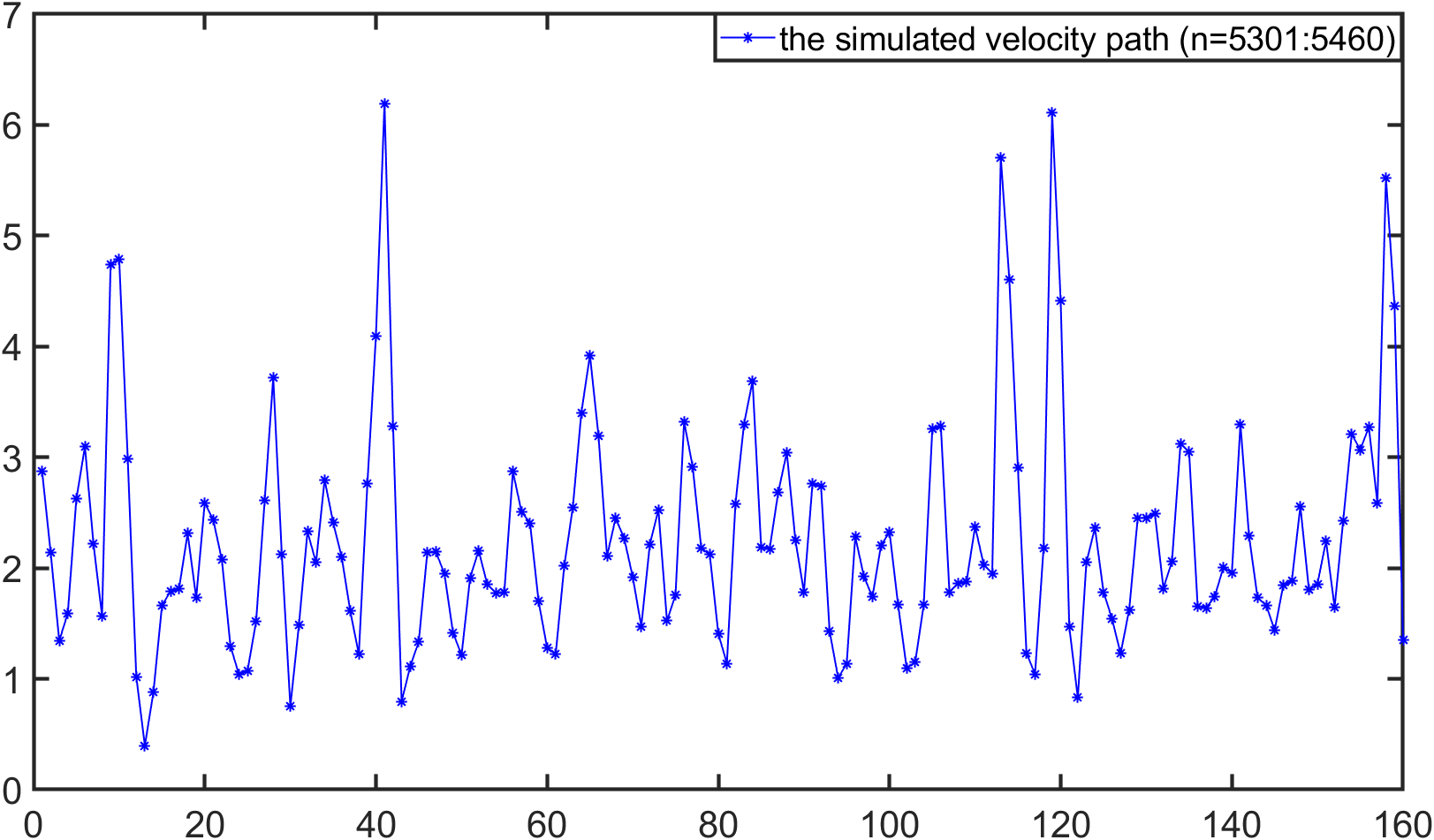}
}
\quad
\subfigure[]{
\includegraphics[width=3.5cm]{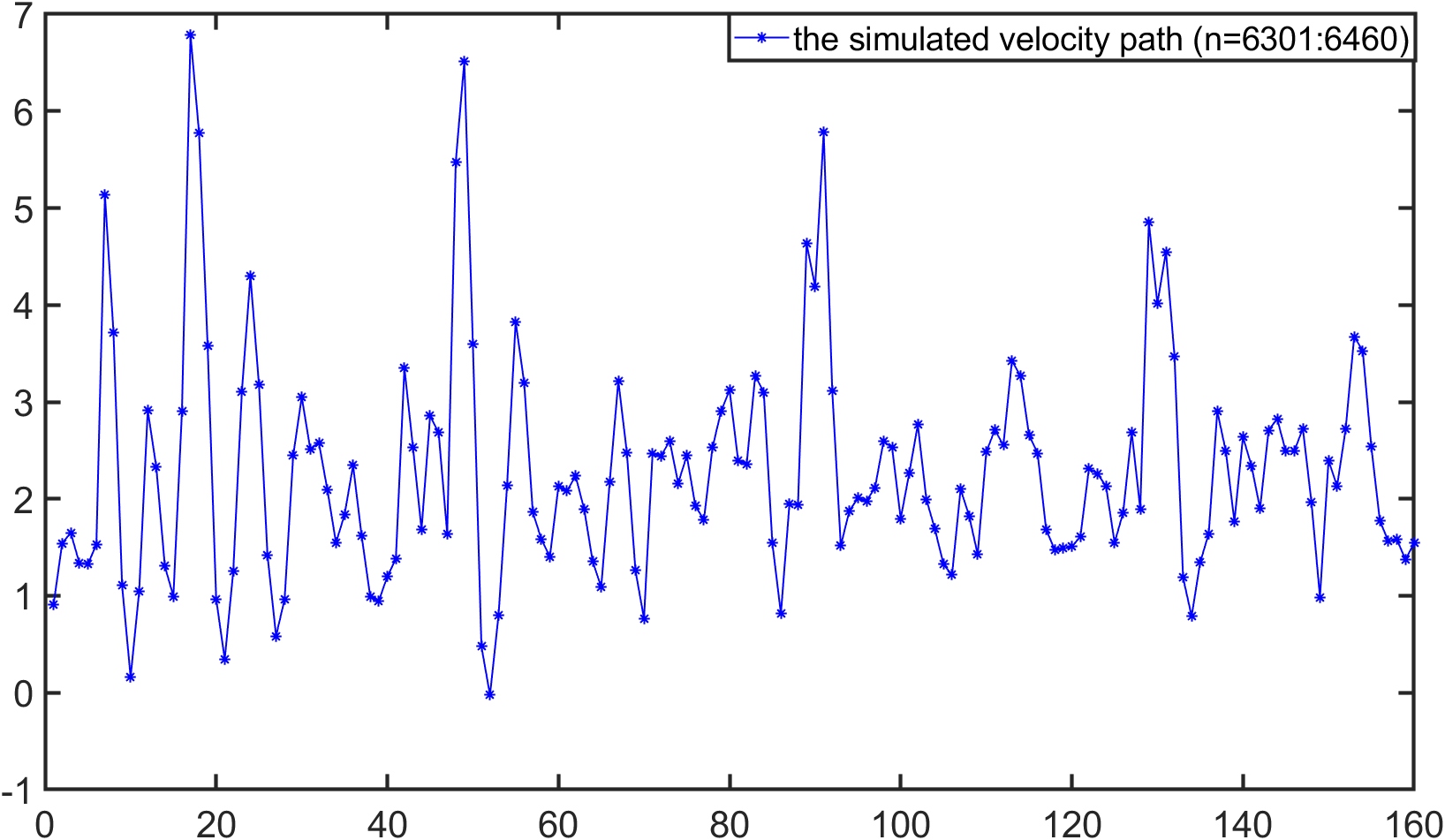}
}

\caption{Long time simulated trajectory of velocity by using the triangular distribution.}\label{longtimetwotr2}
\end{figure}

\noindent{\bf Initial value independence:}
We randomly select several initial values in the algorithm (\ref{newmain}) under the  superposition of two triangular distributions reproducing  expectation and variance, and the simulated results are shown in Figure \ref{initialtwotr2}. It can be seen that the pattern of simulated trajectory of the velocity $v$ is independent of the initial values.
\noindent{\bf Long time behavior:}
Based on the algorithm (\ref{newmain}), we get the simulated velocities after 6500 iterations by using the superposition of two triangular distributions reproducing  expectation and variance. Figure \ref{longtimetwotr2} shows trajectories on several specific time periods.

The behavior of the simulated velocity shows that the pattern will be preserved  in equilibrium state for long time exhibiting the chaotic phenomena, sudden accelerations and the continual random oscillations.

\section{Conclusion}
In this paper, we derive a soft Lasso's approach to learn the soft dynamical equation for the physical mechanical relationship of the motion of a ball falling in non-Newtonian fluids. Under a new concept of soft matching for mass points, several empirical distributions of velocity $v$ are constructed and used in a discrete iterative algorithm which combines the Newton-Stokes term and the soft Lasso's term. The theory is validated by numerical examples, which exhibit the chaotic phenomena, sudden accelerations and the continual random oscillations. Furthermore, the pattern of the motion is independent of initial values and is preserved for long time.
\section*{Acknowledgments}
This work was supported by National Natural Science Foundation of China (12061160462, 11631015).
The authors thank the anonymous referees and the editors for their helpful comments on the manuscript.


\bibliographystyle{elsarticle-num}
\bibliography{sample}

\end{document}